\documentclass{vldb}
\usepackage{graphicx}
\usepackage{hyperref}
\usepackage{balance}  

\usepackage{subfigure}
\usepackage{multirow}
\usepackage{url}
\usepackage{balance}
\usepackage{color}
\usepackage{xspace}
\usepackage{times}

\usepackage{amsmath}

\usepackage{array}
\makeatletter
\newcommand{\thline}{%
    \noalign {\ifnum 0=`}\fi \hrule height 1pt
    \futurelet \reserved@a \@xhline
}
\newcolumntype{"}{@{\hskip\tabcolsep\vrule width 1pt\hskip\tabcolsep}}
\makeatother

\newcommand{\eat}[1]{}

\def\papernumber #1 raised #2 {
  \vspace{-#2}
  \vbox to 0pt{\framebox{\bf Paper Number: #1}}
}

\newcommand{\smgap}{-5pt}

\newtheorem{definition}{Definition}

\newtheorem{example}{Example}

\vldbTitle{TableQnA: Answering List-Intent Queries With Web Tables}
\vldbAuthors{Kaushik Chakrabarti, Zhimin Chen, Siamak Shakeri, Guihong Cao and Surajit Chaudhuri}
\vldbDOI{https://doi.org/TBD}
\vldbVolume{12}
\vldbNumber{xxx}
\vldbYear{2019}

\begin{document}


\title{TableQnA: Answering List-Intent Queries With Web Tables}



%
%
%
%

\numberofauthors{5} 

\author{
%
%
\alignauthor
Kaushik Chakrabarti \\
       \affaddr{Microsoft Research}\\
       \email{kaushik@microsoft.com}
\alignauthor
Zhimin Chen \\
       \affaddr{Microsoft Research}\\
       \email{zmchen@microsoft.com}\\
\alignauthor Siamak Shakeri \titlenote{Work done while at Microsoft (Bing).}\\
       \affaddr{Amazon}\\
       \email{siamaksha1@gmail.com}\\
\and  
\alignauthor Guihong Cao\\
       \affaddr{Microsoft (Bing)}\\
       \email{gucao@microsoft.com}\\
\alignauthor Surajit Chaudhuri \\
       \affaddr{Microsoft Research}\\
       \email{surajitc@microsoft.com}\\
}

\maketitle

\begin{abstract}
The web contains a vast corpus of HTML tables.
They can be used to provide direct answers to many web queries. We focus on answering two classes of queries with those tables: those seeking
lists of entities (e.g., `cities in california') and those seeking superlative entities (e.g., `largest city in california').
The main challenge is to achieve high precision with significant coverage.
Existing approaches train machine learning models to select the answer from the candidates; they rely on textual match features between
the query and the content of the table along with features capturing table quality/importance. 
These features alone are inadequate for achieving the above goals.
Our main insight is that we can improve precision by (i) first extracting intent (structured information) from the query for the above query classes  and (ii) then
performing structure-aware matching (instead of just textual matching) between the extracted intent  
and the candidates to select the answer.
We model (i) as a sequence tagging task. We
leverage state-of-the-art deep neural network models with word embeddings.
The model requires large scale training data which is expensive to obtain via manual labeling;
we therefore develop a novel method to automatically generate the training data.
For (ii), we develop novel features to compute structure-aware match
and train a machine learning model.
Our experiments on real-life web search queries show that (i) our intent extractor for list and superlative intent queries has significantly higher precision and coverage
compared with baseline approaches and (ii) our table answer selector significantly outperforms
the state-of-the-art baseline approach. This technology
has been used in production by Microsoft's Bing search engine since 2016.
\end{abstract}

\section{Introduction} \label{sec:intro}

\begin{figure}[t]
\vspace{-0.4cm}
\begin{center}
\includegraphics[width=3.2in]{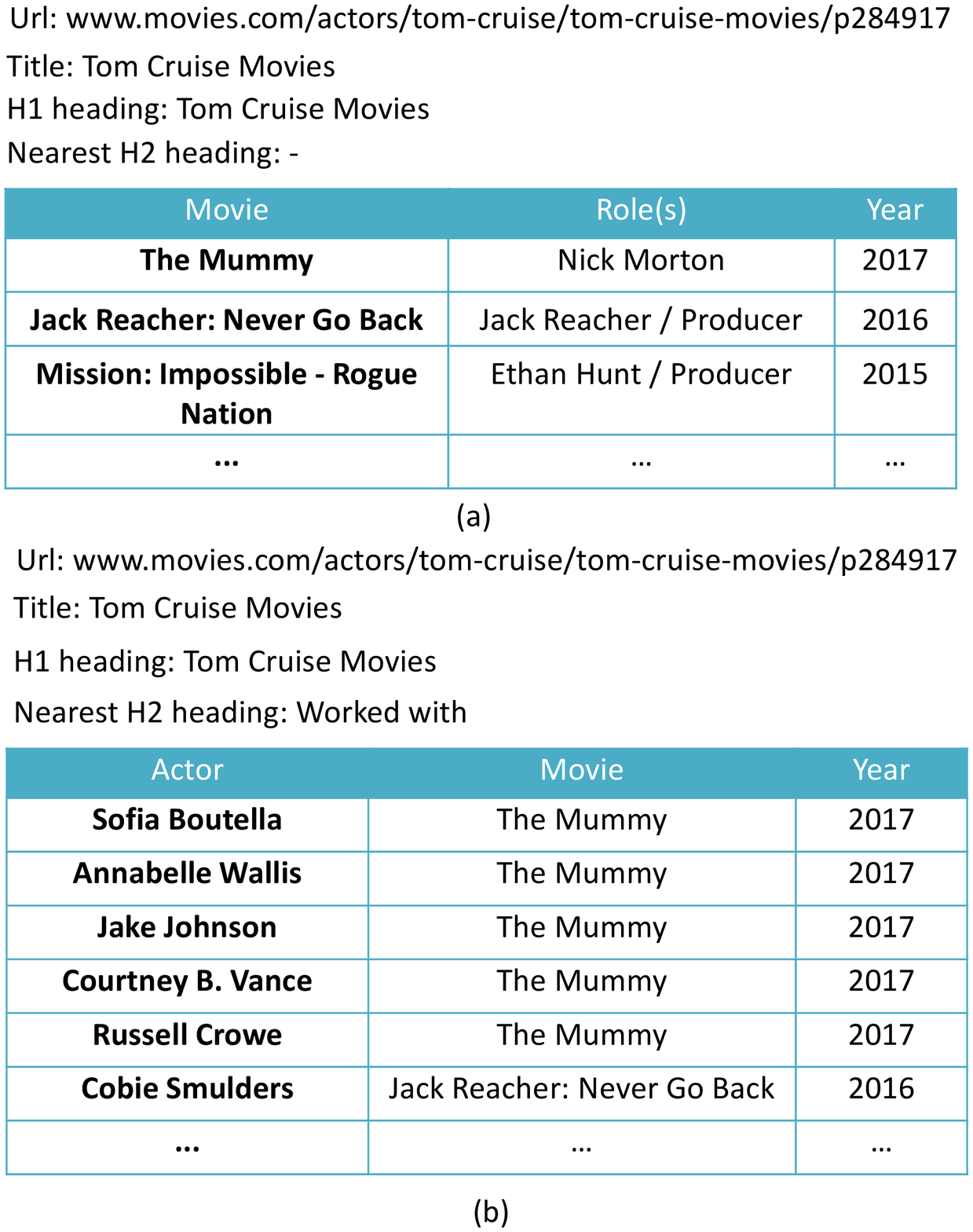}
\vspace{-0.3in}
\caption{\small \label{fig:webtableexample} Web Table Examples}
\end{center}
\vspace{-0.3in}
\end{figure}

The web contains a vast corpus of HTML tables.
We focus on one class
of HTML tables called ``relational tables'' \cite{cafarella:vldb08,cafarella:vldb09,venetis:vldb11}.
Such a table contains a set of entities and their values
on various
attributes,
each row corresponding to a distinct entity and each column
corresponding to a distinct attribute.
Figure \ref{fig:webtableexample} shows two examples of such tables.
For the rest of this paper,
we refer to such tables as simply web tables.
Note that tables that contain information about a single entity (e.g., infobox tables in Wikipedia)
do not fall in this class.

These tables contain a large amount of structured information on a wide variety of topics and
can be used to provide direct answers to many web  queries \cite{facto:www11,sun:www2016}.
%
We focus on answering 
two important classes of queries using those tables: \\
\noindent $\bullet$ \emph{List-of-entity query} (\emph{list-intent query} in short): It
seeks a list of entities of a specific type (referred to as \emph{sought entity type}) that satisfies certain criteria (referred to as \emph{filtering criteria}). 
Both the sought entity type and the filtering criteria are specified in the query string. An example is `coastal cities in california'; the sought entity type
is `city' and the filtering criteria are `coastal' and `in California'.
\\
\noindent $\bullet$ \emph{Superlative-entity query} (\emph{superlative query} in short): It seeks
one or more entities of a specific type that
not only satisfies certain criteria but is also superlative in terms of a \emph{ranking criterion}. The sought entity type,
filtering criteria and ranking criterion are specified in the query string.
An example is `largest city in california'; the sought entity type is `city', the filtering criterion is `in california’ and the
ranking criterion is `largest'.
\\
Our analysis of Bing search logs show
that more than 10\% of Bing queries fall in these classes. We consider these two classes together
since they can be answered using similar approaches. There are some queries that simultaneously satisfy both criteria, e.g.,
`largest cities in california'. We consider them to be list-intent queries for simplicity.
In spite of their importance, there is little work in the research literature on answering such queries.
This paper attempts to fill that gap.

Since relational tables contain sets of entities and those are often superlative entities \footnote{\small{There are many tables about superlative sets
of entities like the largest cities in united states, highest mountains in asia, cities with highest poverty, etc.}. They are suitable for answering superlative queries.}, 
they are naturally suitable for answering such queries.
For example, the table in Figure \ref{fig:webtableexample}(a) is a good direct answer for the list query `tom cruise movies'.
We refer to such answers as \emph{``table answers''}.

\noindent \textbf{Prior work:}
There is extensive work on question answering (QA) for fact lookup queries.
They typically do so by leveraging
a knowledge graph \cite{berant:emnlp2013,yih:acl2015,fader:kdd2014,unger:www2012,yahya:emnlp2012}
or text passages extracted from web pages \cite{pascabook:2003,jurafskybook:qachapter,brill:emnlp2002,lin:tois2007}.
There is also work on using web tables to answer such queries \cite{facto:www11,sun:www2016}.
However, these techniques are not applicable for answering list and superlative queries.

There also exists a rich body of work on table search \cite{cafarella:vldb08,venetis:vldb11,bhagavatula:idea13}.
Given a keyword query, the goal is to return a ranked
list of web tables
relevant to the query.
Prior work first identifies a pool of \emph{``candidate tables''}, typically by sending the query to a web search engine.
Subsequently, it develops features and trains a machine learning (ML) model to rank those candidates.
The features include
textual match between the query and the content inside and around the table in the containing page.
They also include features capturing the quality of the table (e.g., fraction of empty cells)
as well as the importance of the table relative to the document
(e.g., fraction of document occupied by the table).

A baseline approach is to use the features developed for table search and train a classifier 
to select the table answer, if one exists, from the candidates.
This approach has two limitations.\\
\noindent (a) \emph{Table not appropriate type of answer}: None of the above features try to understand/classify
the intent of the query. So, even if there exists a ideal table based on above features, a
table may not be the appropriate \emph{type} of answer
for the query.
Consider the query `michael phelps'.
Most users are looking for important information about the person like his profession, age,
key accomplishments and so on.
There exists a table in his Wikipedia page containing all his world records and exact timing. This table
has perfect textual match and is of high quality; we still should not return that table answer.
It is better to return the entity information panel (typically shown on the right in search engines) containing the above information.
\\
\noindent (b) \emph{Features not adequate}: Even when table is the ideal answer type, the above features may not be adequate
to identify the right table.
Consider the query `tom cruise movies'; a table is the ideal answer type for it. The
two tables shown in Figure \ref{fig:webtableexample} (both from
the same url)
are in the candidate set. The first one contains the movies that Tom Cruise acted in/produced while the second contains
his co-actors in various movies.
Both tables have perfect textual match (with the title and H1 heading),
are of high quality and are important relative to the document. However, the first table is a right answer while
the second one is not.
Since the above features cannot distinguish between the two tables, it 
might pick one the two tables arbitrarily as the answer leading to lower precision.

\noindent \textbf{Proposed QA Framework:}
%
%
%
To improve precision, we need to address limitations (a) and (b).
We propose a 2-step QA framework to address those limitations: \\
\noindent (i) \emph{Intent extraction/classification for list and superlative intent query}: If the query has list or superlative intent, 
we extract the specific intent from it. Specifically, we extract 
the \emph{type of entities} sought by it and the \emph{filtering criteria}.
If the query does not have the above intents, the intent extractor produces a null output.
Thus, the intent extractor also implicitly classifies whether the query  
has list or superlative intent or not. 
We attempt to answer it (i.e., pass it to the next step) only if the extracted intent is non-null.
Although there is rich body of work
on query intent classification into target categories (e.g., Wikipedia categories), we
are not aware of any work on intent extraction/classification for list/superlative intent \cite{queryclass:www09,queryclass:tois06}.
This step
addresses limitation (a) as
relational tables is a good type of answer for such queries.
\\
\noindent (iii) \emph{Structure-aware matching}: 
The intent extracted from the query is more structured compared to the sequence of words in the original query.
The table is also structured.
We take advantage of the structure on both sides to ensure a stricter and more fine-grained match between the query and the selected table. 
For example, we require the type of entities occurring in the selected table to match
with the type of entities sought by the query.
We refer to it as \emph{``structure-aware matching''}.
Specifically, we introduce novel features for the ML model that computes such matches.
The example below shows how these new features addresses limitation (b).
\\
We have built a QA system, called \textsc{TableQnA}, based on the above framework.

\begin{example} \label{ex:sam}
We assume that there is a dictionary
\{{\sf{city}}, {\sf{film}}, {\sf{school}}, $\ldots$\}
of entity types we want to detect in list and superlative intent queries;
it can be obtained from an existing ontology like Freebase.
For simplicity, each element in the dictionary represents both the semantic entity type as well as 
the string users use in such queries to specify the type of entities they are seeking.
We refer to them as type name strings.
In practice, this does not need to be a 1:1 relationship, i.e.,
an entity type can have multiple type name strings associated with it.
We refer to this as the entity type dictionary or simply type dictionary.

Consider the query `tom cruise movies'.
\textsc{TableQnA}'s intent extractor determines that it is a list intent query and 
the  token `movies' implies the user is seeking entities of type {\sf{film}}.
We refer to the latter as ``sought entity type'' (SET in short) and
the former as ``phrase specifying sought entity type'' (PST in short).
Since the output is non-null, it attempts to answer the query.

Recall that each row in a
relational table corresponds to an entity; there is
typically a single column that contains the names of those entities \cite{venetis:vldb11}.
This is referred to as the subject column of the table.
For example, in the table in Figure \ref{fig:webtableexample}(a), each row
corresponds to a movie entity. The name of the movie entities are in the leftmost column (highlighted in bold), hence
that is the subject column.
In the table in Figure \ref{fig:webtableexample}(b), each row corresponds to
a co-actor. The name of the co-actors are the leftmost column (highlighted in bold), so that is the subject column.
We assume the subject column has been identified for each table; we provide further details
on subject column identification later \cite{venetis:vldb11}.

One of the structure-aware match features used by the table answer selector is the match between the SET/PST
and the type of entities in the subject column of each candidate table.
The type of entities in the subject column of the first table is {\sf{film}}
while that of the second table is {\sf{actor}};
the first one
matches with the SET/PST
while the second one does not.
As a result, table answer selector correctly selects the first table as the table answer.
This shows structure-aware matching can determine the correct answer where the above features (specifically, textual match features) cannot.
\end{example}

\begin{figure}[t]
\vspace{-0.4cm}
\begin{center}
\includegraphics[width=3.2in]{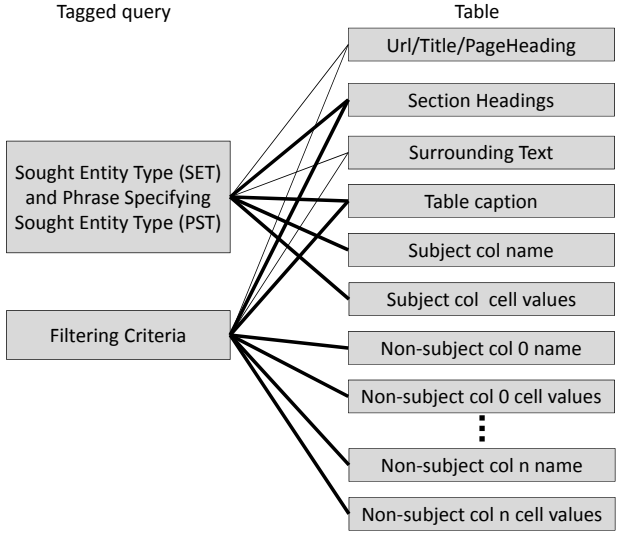}
\caption{\small \label{fig:structMatch} Structure aware match features. Each line represents a match between the extracted query intent and a ``field'' of the table
and is a feature for the ML model. Thicker lines indicate more important features.}
\end{center}
\vspace{-0.3in}
\end{figure}

\noindent \textbf{Key Technical Challenges:}
There are several technical challenges in building such a system.
First, detecting the SET/PST in a list intent query with high precision and recall
is hard.
We consider a simple baseline approach based on exact lookup in the type dictionary.
If a query (exactly) contains a type name string in the type dictionary in plural form, we infer it is a list intent query seeking the
corresponding type of entities. If it contains a type name string in singular form and starts with a superlative keyword (checked using another dictionary, details in Section \ref{sec:tagging}), it is a superlative query. If neither conditions are satisfied, we conclude
it is neither list nor superlative query and hence produce null output.
We refer to this approach as ``type dictionary lookup'' (TDL in short).
TDL suffers from both low precision and recall.
\\
\noindent $\bullet$ \emph{Low precision}: Occurrence of a type name string in plural form does not guarantee
that it is a list or superlative query. For example,
`joan rivers' is not seeking a list of rivers or `physical education in schools' is not seeking for a list of schools.
This leads to low precision.
\\
\noindent $\bullet$ \emph{Low recall}:  
TDL can detect a list intent query only when the PST is exactly identical to 
the type name string in the dictionary.
It will miss any query which uses a different PST to specify the entity type.
For example, since the type name string in the dictionary for the entity type {\sf{film}} is `film',
TDL will miss the queries `tom cruise movies', `tom cruise flicks' and `tom cruise filmography'.
This leads to poor recall.
This problem can be somewhat mitigated by adding multiple type name strings for each entity type to capture
the different ways it can be specified in the query.
However, this is not a complete solution
as it is not feasible to manually add
all possible strings for each entity type.

The second challenge is developing features for structure-aware matching between the extracted query intent and any candidate table.
Figure \ref{fig:structMatch} shows the features along with their importance.
For example, one challenge is to compute match between
the SET/PST
and the type of entities in the subject column.
This would be easy if we knew the type of entities (from the entity type dictionary in Example \ref{ex:sam}) in the subject columns of tables.
In some cases, the name of the subject column
indicates the type of entities in it; there we can simply match with the subject column name.
However, this is often not the case.
For example, the name of the subject column could simply be `Name' or it may even not have any name.
Computing the match accurately is a challenge in these cases.

%

\noindent \textbf{Contributions:} Our technical contributions can be summarized as follows:\\
\noindent $\bullet$ We introduce the framework for answering list and superlative intent queries
with web tables (Section 2). To the best of knowledge, this is the first
paper to focus on answering these classes of queries.
 
\noindent $\bullet$ We model the intent extraction task for list and superlative intent queries
as a sequence tagging task.
We overcome the limitations of TDL by leveraging state-of-the-art deep neural network models, specifically a bidirectional Long
Short-Term Memory (BiLSTM) model with a softmax layer.
One of the main challenges is that the model requires large amount of training data;
manually labeling such data is expensive.
We develop a novel method of automatically generating large-scale training data to address the problem.
Our training data generation algorithm addresses the precision issue of TDL
while the deep learning methods (via modeling of semantic similarity) improves recall (Section 3).

\noindent $\bullet$ We develop novel features for structure-aware matching between the extracted query intent and any candidate table.
We use these features along with textual match and table quality/importance features
and train a classifier to identify the table answer (Section 4).

\noindent $\bullet$ We perform extensive experiments on real-life search queries
in Bing search engine and real web tables extracted from web pages (Section 6).
Our experiments show that (a) our deep model-based query intent extractor has significantly higher precision and coverage
over TDL and other baseline approaches and (b) our table answer selector significantly outperforms
state-of-the-art baselines in terms of precision and recall.

\section{System Architecture and Problem Statement} \label{sec:problem}

We first describe the system architecture. We then formally define the
two technical problems we study in this paper: (i) intent extraction for list and superlative intent queries
and (ii) table answer selection.

\begin{figure}[t]
\vspace{-0.4cm}
\begin{center}
\includegraphics[width=3.5in]{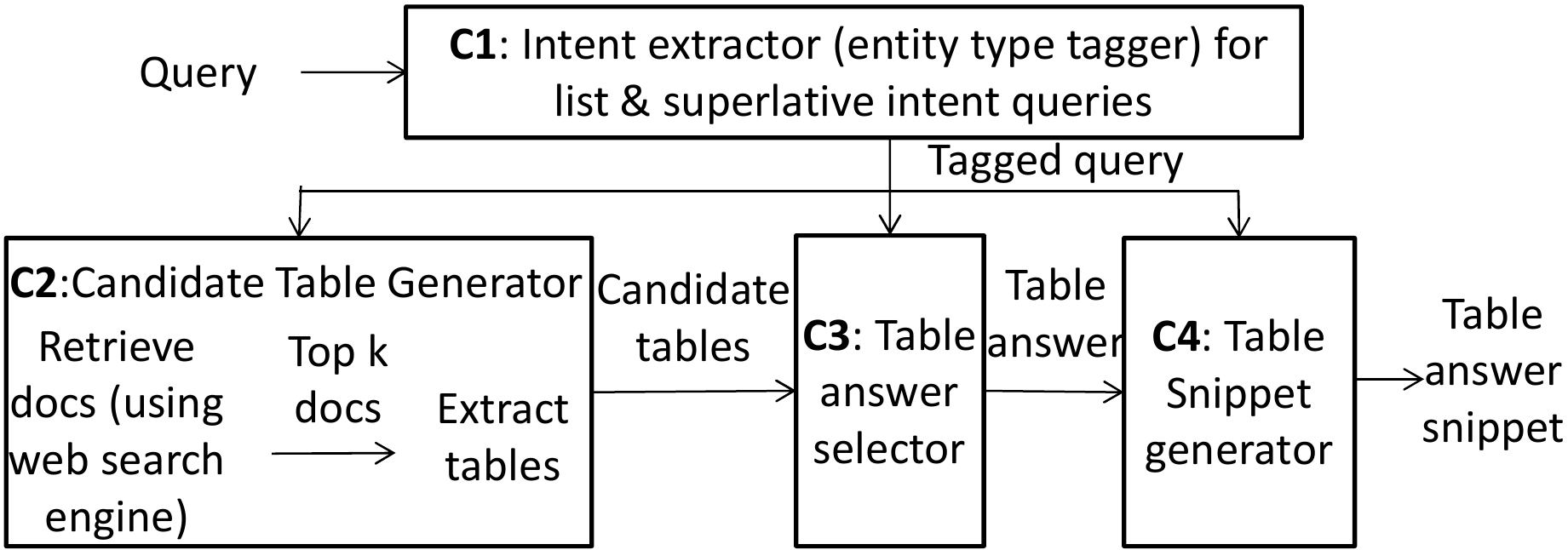}
\caption{\small \label{fig:archi} System architecture}
\end{center}
\vspace{-0.6cm}
\end{figure}

\subsection{System Architecture}
Figure \ref{fig:archi} shows the architecture of the \textsc{TableQnA} system.
It consists of 4 components:
\\
\noindent \textbf{C1. Intent extractor/classifier for list and superlative intent queries}: This component takes the incoming user query as input and
extracts intent from it if it has list or superlative intent. If it does not have the above intents, it produces null output.
\textsc{TableQnA} attempts to answer the query (i.e., passes it to the next component) only if the extracted intent is non-null.
This component therefore has dual purpose. First, it classifies whether the query has list or superlative intent or not.
Second, it extracts the intent (type of entities being sought, filtering criteria) for such queries 
so that table answer selector can leverage it.
Since the main extracted information is the type of entities being sought (the filtering criteria being a byproduct),
we henceforth (for concreteness) refer to this component as \emph{entity type tagger} and its output as \emph{tagged query}.
\\
\noindent \textbf{C2. Candidate table generator}: This component takes the tagged query passed by the first component as input
and generates the \emph{candidate tables} among which the table answer is selected.
It does so by sending the query to a web search engine,
retrieving the top $k$ ($k \approx 5-10$ ) documents and extracting the tables occurring in them.
This is similar to QA systems that return text passages as answers \cite{pascabook:2003,jurafskybook:qachapter}.
We adopt the techniques previously proposed to extract and classify relational tables \cite{cafarella:relWT,wang:www02}.
For each table, we extract the following information: \\
\noindent $\bullet$ url, title and h1 heading of the document \\
\noindent $\bullet$ heading of the section/subsection the table belongs to (typically, h2/h3/h4 headings) \\
\noindent $\bullet$ text immediately preceding the table (referred to as surrounding text) \\
\noindent $\bullet$ caption (content of $<$caption$>$ tag) and header/footer rows if they exist \\
\noindent $\bullet$ column name of the subject column. As discussed in Example \ref{ex:sam},
each row in a
relational table corresponds to an entity and there is
typically one column, referred to as the \emph{subject column}, that contains the names of those entities \cite{venetis:vldb11}.
We identify the subject column of the table using techniques similar to \cite{venetis:vldb11}. \\
\noindent $\bullet$ cell values of the subject column \\
\noindent $\bullet$ column names of the non-subject columns and \\
\noindent $\bullet$ cell values of the non-subject columns. \\
\\
\noindent \textbf{C3. Table answer selector}:
This component takes the tagged query and the candidate tables as input and
determines whether any of those tables provides answer to the query. If yes, it identifies that table.
\\
\noindent \textbf{C4. Table snippet generator}: Given the tagged query and the selected table answer,
this component computes a ``snippet'' of the table to be displayed on the search engine result page.
\\

While in some cases the entire table is the answer, only a part of the table is the answer in other cases.
For example, for the query `tom cruise movies', the entire table in Figure \ref{fig:webtableexample}(a) is the answer.
On the other hand, for the query `2017 tom cruise movies', only
only the first row of that table is the answer.
\emph{In both cases, the task of table answer selection is to determine the table that provides the answer}.
Selecting the best set of rows to be displayed in the former case and
the part of the table that answers the query in the latter case is the task of the snippet generator.
We separate the problem in these two steps because there are exponential number of possible table subsets,
hence it is not feasible to compute the subset answer in one step.

We focus on components C1 and C3 in this paper. We formally define the two problems in the next two subsections
and describe our approaches in detail in Section 3 and 4.
Component C2 adopts existing approaches, so we do not discuss it further.
Component C4 is novel but is not the focus on this paper;
we briefly describe it in Section \ref{sec:snippet}.

\begin{table}[t]
\begin{center}
\small{
\vspace{-0.1in}
\begin{tabular}{|p{1.7cm}|p{1.7cm}|p{1.7cm}|p{1.7cm}|}
\hline
{\sf{city}} & {\sf{country}} & {\sf{park}} & {\sf{church}} \\ \hline
{\sf{town}} & {\sf{postal code}} & {\sf{ski area}} & {\sf{area code}}\\ \hline
{\sf{school}} & {\sf{actor}} & {\sf{film}} & {\sf{aquarium}} \\ \hline
{\sf{symbol}} & {\sf{drug}} & {\sf{hospital}} & {\sf{county}} \\ \hline
{\sf{camera}} & {\sf{phone}} & {\sf{computer}} & {\sf{bank}}\\ \hline
{\sf{conflict}} & {\sf{album}} & {\sf{artist}} & {\sf{politician}}\\ \hline
{\sf{composer}} & {\sf{concert}} & {\sf{band}} & {\sf{mountain}}\\ \hline
{\sf{song}} & {\sf{king}} & {\sf{athlete}}  & {\sf{river}}\\ \hline
{\sf{stadium}} & {\sf{golf course}} & {\sf{race}} & {\sf{volcano}}\\ \hline
{\sf{coach}} & {\sf{team}} & {\sf{bridge}} & {\sf{pond}}\\ \hline
{\sf{port}} & {\sf{road}} & {\sf{trail}} & {\sf{island}}\\ \hline
{\sf{attraction}} & {\sf{painter}} & {\sf{building}} & {\sf{glacier}}  \\ \hline
{\sf{skyscraper}} & {\sf{tower}} & {\sf{galaxy}} & {\sf{geyser}} \\ \hline
{\sf{planet}} & {\sf{vehicle}} & {\sf{airport}} & {\sf{desert}} \\ \hline
{\sf{animal}} & {\sf{flower}} & {\sf{plant}} & {\sf{beach}}\\ \hline
{\sf{boat}} & {\sf{author}} & {\sf{book}} & {\sf{cheese}}\\ \hline
{\sf{station}} & {\sf{company}} & {\sf{celebrity}} & {\sf{drink}}\\ \hline
\end{tabular}
}
\end{center}
\caption{Entity type dictionary in \textsc{TableQnA}}
\label{tab:td}
\vspace{-0.2in}
\end{table}

\subsection{Entity Type Tagging Problem}

We formally define the entity type tagging problem for list and superlative intent queries.
We build two separate taggers: one for list-intent queries
and the other for superlative queries. 
We first define the entity tagging problem for list-intent queries.

\begin{definition}[Entity Type Tagging Problem]
Let $\mathcal T$ denote the type dictionary, i.e., the pre-determined set of entity types to be detected in queries.
Consider a query $Q$. Let $Q.toks$ denote the sequence of word tokens in $Q$.
If $Q$ is a list intent query, the entity type tagger returns $\langle Q.pst, Q.type, Q.lscore \rangle$
where $Q.pst$ denotes the subsequence of $Q.toks$
that specify the entity type being sought, $Q.type \in \mathcal T$ the entity type being sought
and $Q.lscore$ the degree of confidence in the tagging result.
Otherwise, it returns null.
Recall that we refer to the latter as ``sought entity type'' (SET in short)
and the former as ``phrase specifying sought entity type'' (PST in short).
We refer to $\langle$ Q, Q.pst, Q.type, Q.lscore $\rangle$
as the tagged list-intent query.
\end{definition}

We refer to the part of the query string to the left of $Q.pst$
as the premodifier (denoted by $Q.premod$) while the part to the right of $Q.pst$ as the postmodifier (denoted by $Q.postmod$).
They typically specify the filtering and ranking criteria.
Both premodifier and postmodifier may be non-empty or one of them may be empty; both are typically not empty at the same time.
Thus, entity type tagger not only extracts the SET
but also the filtering/ranking criteria as a byproduct.

The tagger has a threshold knob $\rho$; it returns
a non-null tagged query only if the confidence $Q.lscore$ exceeds $\rho$.
The choice of $\rho$ allows \textsc{TableQnA} to obtain the desired precision and recall tradeoff.

Entity type tagging for superlative query tagging is almost identical to the one for list intent queries
except that it applies to superlative queries and it returns $Q.sscore$ 
(the degree of confidence in the tagging result) instead of $Q.lscore$.

\begin{example}
Consider the query `tom cruise movies'.
Let $\mathcal T =$ \{{\sf{city}}, {\sf{film}}, {\sf{school}}\}.
The entity type tagger 
returns $\langle `movies', {\sf{film}}, 1.0 \rangle$
indicating that it is a list-intent query and the token `movies' specifies the entity type being sought
is {\sf{film}}.
The premodifier is `tom cruise' and the postmodifier is empty.
For the superlative query `largest city in california', the entity type tagger
returns $\langle `city', {\sf{city}}, 1.0 \rangle$.
\end{example}

\noindent \textbf{Type Dictionary}: Entity type tagging assumes the presence of 
a dictionary $\mathcal T$ of entity types. We manually curate this dictionary by analyzing web search logs; we include 
the common entity types sought in list intent
and superlative queries.
In \textsc{TableQnA}, we have 68 such types; the type dictionary is shown in Table \ref{tab:td}.
The type name strings are typically single word but they can be a multi-word phrase (e.g., ``golf course'', ``area code'').

The type dictionary contains only those strings whose mention in the query \emph{unambiguously} indicate that
it is seeking entities of that type.
Consider the string ``place''.
In the query `best places to live in wa', it refers to
the type {\sf{city}} while in `best places to work in wa', it refers to
the type {\sf{company}}. Hence, it is ambiguous.
We do not include such strings since including them will lead to wrong tagging.

\begin{table}
\begin{center}
\vspace{-0.1in}
\begin{tabular}{|p{0.05in}p{3.0in}|}
\hline
1 & For each candidate table $T \in \mathcal T$, compute $F(Q,T)$ by invoking the table answer classifier. \\
2 & Pick the candidate table $T_{best} = argmax_{T \in \mathcal T} F(Q,T)$ with the highest classifier score \\
3 & If $F(Q,T_{best}) > \theta$, return $T_{best}$, else return $\{\}$. \\
\hline
\end{tabular}
\caption{Selecting table answer based on table classification results}
\label{tab:tas}
\end{center}
\vspace{-0.2in}
\end{table}

\subsection{Table Answer Selection Problem}
We formally define second technical problem we address in this paper: table answer selection.
\begin{definition}[Table Answer Selection Problem]
Given a tagged list intent (or superlative) query $\langle$ Q, Q.pst, Q.type, Q.lscore $\rangle$ (or $\langle$ Q, Q.pst, Q.type, Q.sscore $\rangle$)
and a pool $\mathcal C$ of candidate tables,
determine whether there exists any candidate table that provides answer to the query.
If yes, return the best such candidate table $T_{best} \in \mathcal T$,
otherwise return $\{\}$.
\end{definition}

Our approach is to first solve the table answer classification problem which is defined as follows:
\begin{definition}[Table Answer Classification Problem]
Given a tagged list intent (or superlative) query $\langle$ Q, Q.pst, Q.type, Q.lscore $\rangle$ (or $\langle$ Q, Q.pst, Q.type, Q.sscore $\rangle$)
and a table $T$, return the score $F(Q,T)$
which represents the degree to which $T$ provides answer for $Q$.
\end{definition}

Subsequently, we perform the table answer selection using the algorithm shown in Table \ref{tab:tas}.
It invokes the table answer classifier for each candidate table; this is feasible
as we typically have a small set of candidate tables. It then
picks the one $T_{best}$ with the highest classifier score.
If $T_{best}$'s score exceeds a threshold $\theta$, we return it as the answer, otherwise we return no answer.
The choice of $\theta$ allows \textsc{TableQnA} to obtain the desired precision and recall trade-off.

\section{Entity Type Tagging} \label{sec:tagging}
In this section, we describe our approaches for entity type tagging for list and superlative intent queries.
We start with baseline approaches
and then present our proposed approach.

\subsection{Baseline Approaches} \label{sec:baseline}
We consider three baseline approaches, viz. TDL, TDL+ER and TDL+ER+DP.
\subsubsection{Type Dictionary Lookup (TDL) Approach}
This is the simple approach described in Section \ref{sec:intro}.
For any query, we first check whether it contains
any type name string in the type dictionary in plural form. If yes, we infer it as a list intent query.
For example, the query `tom cruise films' contains the type name string `film' in plural form,
hence we infer it is a list intent query seeking entities of type {\sf{film}}.
If not, we check whether it contains any type name string in singular form.
If found, we further check whether it starts with a superlative keyword; we
do this by using a dictionary containing all superlative keywords (e.g., ``largest'', ``smallest'', ``highest'', etc.).
If both conditions are satisfied, we infer it as a superlative query.
For example, the query `largest city in california' satisfies both conditions,
so we infer it is a superlative query seeking entity of type {\sf{city}}.
If neither conditions are satisfied, we conclude that the query has neither list nor superlative intent
and hence produce null output.

For simplicity, we henceforth focus our discussion on list intent queries;
our techniques apply to superlative queries as well. We conduct experiments
for both types on queries in Section \ref{sec:expt}.

\eat{
\begin{table}[t]
\begin{center}
\vspace{-0.1in}
\begin{tabular}{|p{1.5cm}|p{5cm}|}
\hline
Type Identifier  & Type Mention Phrases (TMPs) \\ \hline
{\sf{cityT}}  & city, metro, town, urban center \\ \hline
{\sf{movieT}} & movie, film, flick  \\ \hline
{\sf{schoolT}} & school, college, university, preschool \\ \hline
{\sf{drugT}} & drug, medicine, medication, pill \\ \hline
{\sf{celebrityT}}   & celebrity, celeb, superstar, mogul  \\ \hline
\end{tabular}
\end{center}
\caption{Part of entity type dictionary in \textsc{TableQnA}}
\label{tab:td}
\vspace{-0.2in}
\end{table}
}

\subsubsection{Type Dictionary Lookup with Entity Removal and Dependency Parsing}
As discussed in Section \ref{sec:intro}, TDL suffers from both low precision and recall.
We first turn our attention to improving precision. We identify 2 main reasons for
false positives and develop solutions for them:\\
\noindent $\bullet$ \emph{Entity names}:
One reason for false positives is that the
PST detected by TDL is not intended to be a PST. Rather, \emph{it is part of an entity name and the intent is seek to information about that entity}.
Consider the query `joan rivers'.
TDL detects the PST `rivers' with SET being {\sf{river}}. However, it is not seeking list of
entities of type {\sf{river}}
but information about the celebrity `Joan Rivers'.
Other examples include `tyra banks', `parks and recreation' and `turner classic movies'.
This is quite common and leads to significant loss of precision.

We can simply remove such queries by looking up the query string in a dictionary of entity names (e.g., constructed
from Wikipedia or Freebase). However, that is not adequate.
Consider the query `joan rivers net worth'. This will not get removed.
We need to make sure that the detected PST is not part of an entity name, i.e.,
\emph{no substring of the query containing the detected PST is an entity name}.
We use the above entity name dictionary to remove such queries.
Note that the premodifier and postmodifier can contain entity names (e.g., `tom cruise movies' contains an entity name in the premodifier
while `cities in california' contains an entity name in the postmodifier); the above technique will not eliminate such queries.

\begin{figure}[t]
\vspace{-0.4cm}
\begin{center}
\includegraphics[width=2.4in]{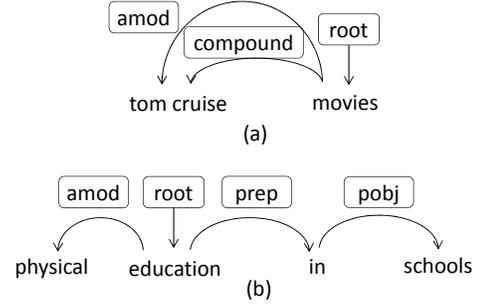}
\vspace{-0.15in}
\caption{\small \label{fig:dp} Dependency parsing examples (a) Query where PST is root word (b) Query where PST is not a root word}
\end{center}
\vspace{-0.3in}
\end{figure}

\noindent $\bullet$ \emph{PST not root word}:
In a valid list intent query, the PST should linguistically be the root word (i.e., linguistically independent of other words)
and all other words should be modifiers (i.e., linguistically dependent on other words).
For example, in the query `tom cruise movies', the PST `movies' is linguistically the root word
as shown in Figure \ref{fig:dp}(a). Hence, it is a valid list intent query.
On the other hand, in the query `physical education in schools', `education' is the root word , not the detected PST `schools'
as shown in Figure \ref{fig:dp}(b). Hence, it is not a valid list intent query.
TDL does not ensure this;
this leads to many false positives.
We address this problem by linguistically parsing the query
and ensuring that the detected PST is the root word;
we refer to it as \emph{``DP root check''} in short. 
It will eliminate false positives like `physical education in schools'
and retain the true positives like `tom cruise movies'.
In \textsc{TableQnA}, we use dependency parsing where each node is a word in the query
and each edge is a linguistic dependency relationship between two words
(as shown in Figure \ref{fig:dp}). In \textsc{TableQnA}, we use the Python package spacy
for dependency parsing.

We refer to the approach that performs only entity removal on TDL output
as \emph{TDL+ER} while the one that performs both entity removal and DP root check as
\emph{TDL+ER+DP}. These approaches improve precision
but does not help with recall. 
We next propose an approach that addresses both limitations.

\subsection{Proposed Approach}

The above approaches have low recall because it can detect a list intent query only when the PST is exactly identical to
the type name string in the dictionary.
It will miss any query which uses a different PST to specify the entity type.
It not feasible to manually add all possible type name strings for each entity type in the dictionary.
Furthermore, we deliberately exclude ambiguous type name strings from the dictionary, this
will also lead to false misses.

\emph{Our main insight is that we can perform entity type tagging
for these missed queries by observing queries correctly tagged queries that have
(i) semantically similar PSTs
and (ii) semantically similar premodifiers and postmodifiers.}
We informally refer to (i) as semantic similarity and (ii) as context similarity.
Consider the false miss `tom cruise movies'.
Now consider the correctly tagged query $\langle$ `tom cruise films,  `films', {\sf{film}} $\rangle$.
Since the token `movies' is similar to the PST `films'
in the semantic space (e.g., using word embeddings \cite{pennington2014glove, mikolovword2vec})
and the premodifier and postmodifiers are identifical
(premodifier is `tom cruise' and postmodifier is empty
in both of them),
we can correctly tag the former query.
Correctly tagged queries
$\langle$ `tom hanks films,  `films', {\sf{film}} $\rangle$
and
$\langle$ `brad pitt films,  `films', {\sf{film}} $\rangle$
will also contribute since the
premodifiers, viz., `tom hanks' and `brad pitt'
are also semantically similar to `tom cruise' as they are all famous male actors.

The above insight works for ambiguous PSTs as well.
We will tag the query `best places to work in wa' with {\sf{company}}
because it has high context similarity to correctly tagged query `best companies to work in wa'
(premodifier is `best' and postmodifier is `to work in wa'). On the other hand, we will
tag the query `best places to live in wa' with {\sf{city}}
because it has high context similarity to correctly tagged query `best cities to live in wa' (premodifier is `best' and postmodifier `to live in wa').

Our entity tagging task can be mapped to the sequence tagging NLP task
\cite{crf:icml01, huang:2015}. Furthermore, for the latter, researchers
have already proposed techniques to learn from labeled sentences
with semantic and context similarities. Specifically, they proposed to use
deep neural networks (DNN), particularly LSTMs, leveraging word embeddings \cite{huang:2015}.
These are the best performing solutions for this task.
So, instead of developing new solutions, we propose to map our task to the sequence tagging task
and leverage the DNN solutions developed for it.

We start by defining the sequence tagging task and discuss the mapping. 
We next describe the sequence tagging DNN model we leverage for entity type tagging.
This model requires large amount of training data; manually labeling such data is expensive.
We conclude this section by discussion how we address this problem.

\subsubsection{Mapping Entity Tagging to Sequence Tagging Task}
The sequence tagging task is defined as follows:
given a sequence of tokens, the task is to assign a categorical label to each token in the sequence.
We informally discuss how we can map the
entity tagging problem defined in Section \ref{sec:problem} to the sequence tagging task;
the exact details can be found in the next two subsections.
Suppose the set of possible labels for the sequence tagging task is $\{ \mathcal T \cup O\}$ where $O$ denotes ``other'' tag;
recall $\mathcal T$ denotes the dictionary of entity types.
We refer to the labels $t \in \mathcal T$ as entity type tags or non-O tags
and the $O$ label as the O-tag.
First consider a list intent query seeking entities of type $t \in \mathcal T$.
The corresponding sequence tagging task is to label the PST tokens with the entity type tag $t \in \mathcal T$ 
and all the other tokens with the `O'-tag.
For a non list intent query, the corresponding sequence tagging task is to label
all tokens with `O'-tag. 


\begin{figure}[t]
\vspace{-0.4cm}
\begin{center}
\includegraphics[width=3.5in]{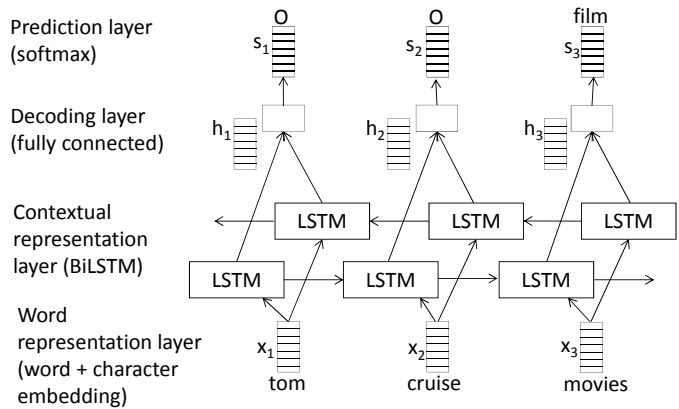}
\vspace{-0.15in}
\caption{\small \label{fig:dnnarchi} Deep model architecture for sequence tagging}
\end{center}
\vspace{-0.3in}
\end{figure}

\subsubsection{Deep Model for Entity Type Tagging} \label{sec:dmarchi}
We adopt the best-performing model for sequence tagging, namely the bidirectional LSTM as proposed in \cite{huang:2015}
and adapt it for our entity type tagging task. Figure \ref{fig:dnnarchi} shows the model architecture.
It consists of 3 layers:
\\
\noindent \emph{1. Word representation layer:} We represent each word $w_i$ in the query in semantic space using word embeddings.
In \textsc{TableQnA}, we use pre-trained word embeddings with 300 dimensions from Glove \cite{pennington2014glove}.
Since many queries contain entity names and those words often do not have pre-trained word vector (out of vocabulary), we
learn embeddings at character level as well and concatenate it with word embeddings to get the final vector $x_i \in R^n$ for each word.
 \\
\noindent \emph{2. Contextual representation layer:} As discussed above, the context of words in
the query play an important role in tagging. For each word $w_i$, we want to construct a vector $h_i \in R^k$
that not just captures its own semantic meaning but also the semantic meanings of all words in its context.
We use an LSTM for this purpose; $h_i$ are the hidden states of each time step.
Since we want to use both left and right contexts, we
use a bi-directional LSTM.
For each time step $i$, we concatenate the hidden states from the two LSTMs to get the final vector $h_i$.
 \\
\noindent \emph{3. Decoding layer:} We use a fully connected neural network to get a vector $s_i$ for each word $w_i$
where each entry corresponds to $w_i$'s score for each tag, i.e., $s = W.h + b$.
We can make the final tag prediction in either local (i.e., for each word independently)
or holistic (i.e., each word's prediction dependent on neighboring predictions) manner. The state-of-the-art approaches perform the latter (using conditional random fields (CRF)).
This is where our task differs from previous sequence tagging tasks like part of speech tagging (POS) or named entity recognition (NER).
In our task, there is \emph{at most one non-O tag
in a query} as \emph{a list intent query typically seeks one type of entities}.
This is not the case in previous sequence tagging tasks like POS and NER where
there are often multiple non-O tags in the sequence.
Hence, we believe the local prediction approach is adequate for our propose.

\begin{figure}[t]
\vspace{-0.4cm}
\begin{center}
\includegraphics[width=2.9in]{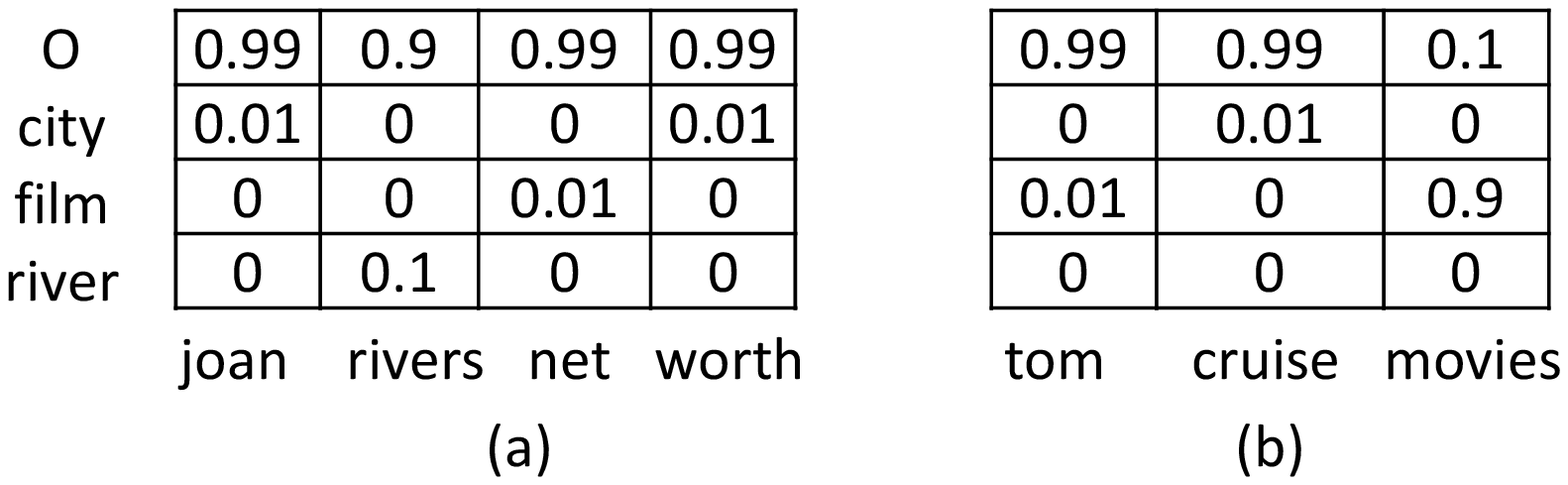}
\vspace{-0.15in}
\caption{\small \label{fig:outputmapping} Mapping sequence tagging deep model output to entity type tagging output}
\end{center}
\vspace{-0.3in}
\end{figure}

The local approach typically uses softmax for final prediction, i.e., for each word $w_i$,
simply pick the tag with the highest score in $s_i$.
However, this does not allow \textsc{TableQnA} to perform precision and recall tradeoff using the threshold $\rho$ as discussed in Section \ref{sec:problem}.
So, we perform the prediction as follows.
For each word $w_i$, we first normalize the scores as follows: $p_i[j] = \frac{e^{s_i[j]}}{\Sigma_j e^{s_i[j]}}$
where $s_i[j]$ and $p_i[j]$ denote the score of word $w_i$ for $j$th tag before and after normalization.
We can interpret the normalized scores $p_i[j]$ as probabilities.
We consider the following cases:\\
\noindent $\bullet$ No word has any entity type tag $t \in \mathcal T$ with (normalized) score greater than $\rho$. 
This means all words have $O$ tags which in turn implies
the query not a list or superlative intent query. Hence, we produce a null output.
\begin{example}
An example of this case is shown in Figure \ref{fig:outputmapping}(a). 
Suppose there are 3 entity types in the type dictionary: {\sf{city}}, {\sf{film}} and {\sf{river}}.
Consider the query `joan rivers net worth'.
The matrix shows the normalized scores $p_i[j]$ of the 4 query words for the 4 tags: the 3 entity type tags 
and the $O$-tag. 
Suppose $\rho = 0.3$.
The scores of all the words for the entity type tags are below $0.3$.
Hence, we produce a null output.
\end{example}
\noindent $\bullet$ There is a continuous sequence of words for which a particular entity type tag $t \in \mathcal T$ has normalized score greater than $\rho$. 
None of the other words have any entity type tag with normalized score higher than $\rho$.
This implies that it is a list intent query with the SET being $t$ and
the contiguous token sequence being the PST. 
We average the normalized scores of the words in the contiguous sequence for the tag $t \in \mathcal T$
to obtain the confidence score $lscore$, i.e., $lscore = avg_{i=a}^{a+b} p_i[j_t]$ where $w_a, w_{a+1}, \ldots, w_{a+b}$ denotes the words in the contiguous sequence
and $j_t$ the index of the entity type tag $t \in  \mathcal T$.
\begin{example}
An example of this case is shown in Figure \ref{fig:outputmapping}(b).
Again, suppose there are the same 3 entity types in the type dictionary.
Consider the query `tom cruise movies'.
The matrix shows the normalized scores $p_i[j]$ of the 3 query words for the 4 tags. Again, suppose $\rho = 0.3$. 
The word `movies' has a score greater than 0.3 for the entity type tag {\sf{film}}, the scores for all the other words for entity type tags are below $0.3$.
Hence, we produce the tagging output $\langle `movies', {\sf{film}}, 0.9 \rangle$.
\end{example}
\noindent $\bullet$ There are non-contiguous words with entity type tag scores greater than $\rho$ or
words have different entity type tags with scores greater than $\rho$. This implies that the query specifies multiple entity types.
This happens rarely in practice; it is also rare in our system since
since all our training data has at most one entity type tag.
If it does happen, we consider it to be a non-list intent query
and produce null output.

\subsubsection{Automatic Large-Scale Training Data Generation} \label{sec:tdgen}
One option is to get the training data via manual labeling. Since DNNs
require large amount of training data (hundreds of thousands), this is expensive.
To address this challenge, we generate the training data automatically.
Since the output of \emph{TDL+ER+DP} has high precision, we
propose to use it as training data.

We take the set of queries from the Bing query click log that has been submitted
at least 100 times (by different users) in the last 2 years. We run \emph{TDL+ER+DP}
on each of them.
We use only the queries with non-null output
as training examples.
We need to generate the labeled example for the sequence tagging model
from the tagged query $\langle$ Q, Q.pst, Q.type $\rangle$ (we omit the $Q.lscore$ for simplicity).
This generation needs to be consistent with our mapping between the two tasks. So, 
we label all the tokens in $Q.pst$ with $Q.type$
and all other tokens with $O$ tag. For example, for the tagged query $\langle$ `tom cruise movies,  `movies', {\sf{film}} $\rangle$,
we generate labeled example:
\begin{align*}
\begin{split}
            & \mbox{tom~~~cruise~~~films} \\
            & \mbox{~O~~~~~~~~O~~~~~~~~{\sf{film}}}
\end{split}
\end{align*}
This will produce a large set of examples for each of the 68 entity types in the type dictionary.

The DNN trained using the above data can detect PSTs
that are semantically similar (but not identical) to the PSTs in the training data, thereby increasing recall.
For example, it will classify the query ‘tom cruise movies'
as a list intent query and detect the PST `movies'.
However, it has an issue. \emph{It does not learn context similarity criterion}, i.e., 
it detects the semantically similar PSTs \emph{irrespective of their contexts}.
As a result, it introduces the entity name (e.g., `joan rivers net worth') and PST-not-root-word 
(e.g., `physical education in schools')
errors back
into the output.
This leads to low precision.

Note that this problem would not arise if the DNN learnt the context similarity criterion (in addition to the PST similarity)
because valid list intent queries do not have contexts like these.
We believe that the DNN is unable to discriminate between 
between valid and invalid contexts
because we provide only valid contexts in the training data.
So, we add queries with invalid contexts in the training data (negative examples).
Recall there are two sources of these: entity names and non-root PST.
So, we add queries removed by entity removal
and DP root-word check into the training set.
These are non list intent examples, so we label all the tokens with 'O' tag:
\begin{align*}
\begin{split}
            & \mbox{joan~~rivers~~net~~worth} \\
            & \mbox{~ O~~~~~O~~~~~~O~~~~~O~~}
\end{split}
\end{align*}

\section{Table Answer Selection}

If the incoming query has non-null entity tagging output,
it is passed to this component.
It also receives a set of candidate tables.
The task of this component is to determine whether any of those tables provides answer to the query.
If yes, it returns the best possible table. Otherwise, it returns an empty result.

As discussed in Section \ref{sec:problem}, this component performs its task by invoking the
table answer classifier. Given a tagged list or superlative query and a candidate table (referred as query-table pair or Q-T pair in short),
the latter classifies whether the table provides answer to the query or not. The rest of the section focuses
on building this classifier.

\begin{table}
\begin{center}
\small{
\vspace{-0.1in}
\begin{tabular}{|p{1.0in}|p{2.2in}|}
\hline
\multicolumn{2}{|l|}{Table Features} \\
\hline
numRows & Number of rows in table \\
numCols & Number of columns in table \\
emptyCellRatio & Fraction of cells in table empty \\
columnNamesPresent & Whether column names present \\
\hline
\multicolumn{2}{|l|}{Table-Page Features} \\
\hline
tableImportance & Inverse of number of tables on page \\
tablePageFraction & Ratio of table size to page size \\
\hline
\multicolumn{2}{|l|}{Page Features} \\
\hline
srRank & Rank of page in search results \\
staticRank & Static rank (PageRank) of page \\
\hline
\multicolumn{2}{|l|}{Query-Table Features} \\
\hline
qInPageTitle & Num query tokens in page title \\
qInTableTitle & Num query tokens in  table caption \\
qInColNames & Num query tokens in  column names \\
qInLeftmostCol & Num query tokens in leftmost column cells \\
qInSecondLeftCol & Num query tokens in second-to-left column cells \\
qInOtherCol & Num query tokens in other column cells \\
qInSurrText & Num query tokens in text preceding table \\
\hline
\end{tabular}
\caption{Previous table search features}
\label{tab:previousfeatures}
}
\end{center}
\vspace{-0.2in}
\end{table}

One option is to simply use the features proposed for table search
in \cite{cafarella:vldb08,venetis:vldb11,bhagavatula:idea13}
and train an ML model to classify Q-T pairs. 
Recall that these features
include
textual match between the query and the content inside and around the table in the containing page.
They also include features capturing the quality of the table (e.g., fraction of empty cells)
as well as the importance of the table relative to the document
(e.g., fraction of document occupied by the table).
Table \ref{tab:previousfeatures} shows the list of all those features \cite{cafarella:vldb08,venetis:vldb11,bhagavatula:idea13}.
We refer to this approach as \emph{BaselineFeatures\_ListQ} (the suffix indicating
that it first performs query intent classification).
BaselineFeatures\_ListQueries is superior to the approach discussed in Section \ref{sec:intro}
(referred to as \emph{BaselineFeatures\_AllQ} since it does not perform any intent classification)
because the former performs 
query intent classification.
Specifically, it addresses the first limitation of the latter (referred to as ``Table not appropriate type of answer'')
discussed in Section \ref{sec:intro}.
However, it does not address the second limitation (referred to as ``Features not adequate'').

We address this limitation by taking advantage of the intent extracted from the query
to compute a stricter and more fine-grained match. We refer to this as
structure-aware matching. Specifically, we introducing novel features for the ML model that computes such matches.
We use them in conjunction with the table search features described above.
We use an off-the-shelf learning algorithm (gradient boosted decision tree) to train the table answer classifier.
We next describe the structure-aware matching features we compute for a given pair 
of query $Q$ and table $T$. 
Note that these features typically have different importance for the classification task.
We simply enumerate the features here; their importance is determined by the ML model based on training data.

\subsection{Match of Sought Entity Type with Subject Column Name} 
The main features of structure-aware match are designed to ensure that the type of entities occurring in the subject column of the table
matches with the type of entities sought by the query (SET).
Recall that
each row in a
relational table corresponds to an entity and there is
typically has one column, referred to as the \emph{subject column}, that contains the names of those entities \cite{venetis:vldb11}.
However, we do not explicitly know the type of entities (from the type dictionary introduced in Example \ref{ex:sam}) in the subject column.
In many cases, the name of the subject column indicates the type of entities in it. So,
we propose \emph{the match between the PST/SET and the subject column name} as a feature. 
We use both the PST and SET to compute the match.
Let $Cont(s,t)$ denote the string containment function, i.e., it returns true when the sequence of tokens in string $s$ contains the tokens in string $t$ as a subsequence 
and false otherwise. 
Let $T.scolname$ denote the name of the subject column of table $T$.
We compute a boolean feature as follows:
\begin{align*}
\begin{split}
            &= 1 \mbox{ if $Cont(T.scolname, Q.pst) \vee Cont(T.scolname, Q.type)$} \\
            &= 0 \mbox{ otherwise}
\end{split}
\end{align*}
We refer to this feature as \emph{SubjectColName\_SET\_Match}.

\subsection{Match of Sought Entity Type with Subject Column Values}
The above feature is not adequate for entity type matching 
since the subject column name often does not indicate the type of entities in it.
It could simply be `Name' or it may not even have a name.
To complement the above feature, we propose to compute \emph{the match between the PST/SET and the cell values in the subject column}.
This is based on the observation that many entities contain the entity type in their names. 
For example, names of most lakes, schools and cities begin or end with the token `Lake', `School' and `City' respectively
(e.g., `Lake Washington', `Interlake High School', `Redwood City').
If a few cell values in the subject column contains the PST/SET, it is likely 
that the type of entities in that column matches the SET.
Let $T.scolcell(i)$ denote the subject column cell value of the $i$th row of table $T$.
We first compute the match of SET/PST with each individual cell values as follows:
\begin{align*}
\begin{split}
            &CellCont(T.scolcell(i),Q)  \\
            &= 1 \mbox{ if $Cont(T.scolcell(i), Q.pst)  \vee Cont(T.scolcell(i), Q.type)$} \\
            &= 0 \mbox{ otherwise}
\end{split}
\end{align*}
We then simply sum up the individual cell value matches to compute the feature: $\Sigma_i CellCont(T.scolcell(i),Q)$.
Higher the number of cell value matches, higher the feature value.
We refer to this feature as \emph{SubjectColValues\_SET\_Match}.

\subsection{Match of Sought Entity Type with Section Headings}
We propose two more features for entity type matching 
to complement the two features above.
One is \emph{the match between the PST/SET and the heading (h2, h3 or h4) of the immediate section/subsection that contains the table}. We refer to this feature as
\emph{SectionHeadings\_SET\_Match}.
A match here strongly indicates that the table contains entities of that type since the table
is typically an important part of the immediately containing section.

A second feature is the match between the PST/SET and 
the headings of all the section/subsections containing the table, including the h1 heading of the document.
We refer to this feature as
\emph{AllHeadings\_SET\_Match}.
This is a weaker feature since the table may not be an important part of the non-immediate containing sections
and/or the document, so a match
here may not indicate that the entities in the table are of that same type. 
Specifically, the boolean features are:
\begin{align*}
\begin{split}
            &= 1 \mbox{ if $Cont(T.sechead, Q.pst) \vee Cont(T.sechead, Q.type)$} \\
            &= 0 \mbox{ otherwise}
\end{split}
\vspace{-1cm}
\end{align*}
and 
\begin{align*}
\vspace{-1cm}
\begin{split}
            &= 1 \mbox{ if $Cont(T.allhead, Q.pst) \vee Cont(T.allhead, Q.type)$} \\
            &= 0 \mbox{ otherwise}
\end{split}
\end{align*}
where $T.allhead$ denotes the concatenation of the headings of all the section/subsections containing the table, including the h1 heading of the document.


\subsection{Match of Premodifier and Postmodifier with Non-Subject Column Values}
The entity type tagging not only outputs the PST and SET but also the 
premodifier and postmodifier.
These typically specify
the filtering and ranking criteria that the entities must satisfy.
There are two scenarios: either the entire table satisfies the 
filtering/ranking criteria or only a subset of entities in the table satisfy the criteria.
For example, for the query `tom cruise movies', the table in Figure \ref{fig:webtableexample}(a) corresponds to the first scenario
while for `2017 tom cruise movies', the same table corresponds to the second scenario.
We require features to ensure match between the filtering criteria 
and the table for both scenarios.

In the first scenario, whether the table satisfies the filtering criteria
can be determined by checking its mention in the page heading and/or section heading. For example, 
the filtering criterion for the query `tom cruise movies' is mentioned in 
the page heading (which is `Tom Cruise Movies') for the table in Figure \ref{fig:webtableexample}(a).
This is already covered
by features proposed by table search (see Table \ref{tab:previousfeatures}).

In the second scenario, we need to look into the attribute values
of the entities to determine whether the table contains any entities
that satisfy the filtering criteria. For example, the filtering criterion `2017' 
appears in the `Year' attribute of table in \ref{fig:webtableexample}(a).
In other words, we need to check whether the 
the tokens specifying filtering criteria (i.e., premodifier and postmodifier tokens)
are present in \emph{the cell values of the non-subject columns} of the table.
Previous approaches do not consider such checks
and will hence miss such table answers since these tokens may not occur anywhere else in the page.
For example, in the above example, the token `2017' does not occur anywhere else in the page.
For each non-subject column $j$, we compute $\Sigma_i CellCont(T.cell(i,j),Q)$ which is computed as:
\begin{align*}
\begin{split}
            CellCont(T.cell(i,j),Q) & = 1 \mbox{ if $Cont(T.cell(i,j), Q.premod)$} \\
                                    & \mbox{~~~~$\vee Cont(T.cell(i,j), Q.postmod)$} \\                                           
                                    &= 0 \mbox{ otherwise}
\end{split}
\end{align*}
To compute the feature, we simply aggregate over all non-subject columns using max.
We also consider both complete and partial matches of the premodifer and postmodifer
as well as number of missing tokens and idf-weighted fractions of missing tokens in premodifier and postmodifier.
We refer to this features together as \emph{PremodPostmod\_Match}.

Note that all the features proposed in this subsection are unique to our system architecture since PST/SET, premodifier
and postmodifier are not available in previous table search systems.


\section{Snippet Generation}
\label{sec:snippet}

Once the table answer selector has selected a table,
the snippet generator chooses a ``good'' $m \times n$ snippet of the table. 
We first define a $m \times n$ snippet.

\begin{definition}[Table Snippet]
Consider a table $T$. Let $\mathcal R_T$ and $\mathcal C_T$ denote the set of data rows and columns of $T$.
A $m \times n$ snippet of $T$ is a table consisting of a subset $SR \subseteq \mathcal R_T$
of $m$ rows and a subset $SC \subseteq \mathcal C_T$
of $n$ columns.
\end{definition}

We display the chosen snippet along with the names of the chosen columns as well the title/h1 heading
and the url of the document. 
The values for $m$ and $n$ are pre-fixed based on
the device screen size (e.g., desktop vs tablet vs phone) and are typically either 3 or 4.

What comprises a good snippet depends on whether the entire table  
or a part of the table is the answer.
\\
\noindent $\bullet$ \textbf{Entire table is the answer}: 
For example, the entire table in Figure \ref{fig:webtableexample}(a) is the answer for the query `tom cruise movies'.
We identify such cases by checking the location of the keyword hits. In such cases, 
there are no keyword hits in the column names or column cell values.
In this scenario, the snippet generator simply chooses the top $m$ rows and the leftmost $n$ columns
while making sure to include the subject column and skipping columns with
too many empty cells/repeated values.
The logic is that the rows and columns in a web table are already
ordered based on a criteria
that the author of the table deems important. For instance, the rows in the table in Figure \ref{fig:webtableexample}(a) are ordered
in descending order of the release year.
This is because the authors thinks that users are more likely interested in Tom Cruise's latest movies rather than his old ones.
\\
\noindent $\bullet$ \textbf{Part of the table is the answer}:
Only some rows and columns contain the answer to the query.
Consider the query `2017 tom cruise movies'. Only the first row is the answer,
so that row must be included in the snippet, preferably at the top.
Again, we identify such cases 
by checking the location of the keyword hits; there are keyword hits in the column names and column cell values, specifically the
cell values of the non-subject columns.
In such cases, we ensure such rows are included in the snippet.
The snippet generator tries to balance between such rows/columns with keyword hits
and the top rows/columns to avoid issues stemming from spurious keyword matches.
We skip the exact algorithms due to space limitations.

\section{Experimental Evaluation} \label{sec:expt}

We present an experimental study of the techniques proposed in this paper. The goal of the study are:\\
\noindent $\bullet$ To assess the importance of list and superlative intent queries  \\
\noindent $\bullet$ To evaluate the quality of DNN-based approach to entity type tagging
and compare it with baseline approaches (TDL, TDL+ER and TDL+ER+DP)\\
\noindent $\bullet$ To evaluate the proposed automatic training data generation approach in
terms of the DNN performance \\
\noindent $\bullet$ To evaluate the impact of the structure-aware matching features on answer quality \\

\begin{table}
\begin{center}
\vspace{-0.1in}
\begin{tabular}{|p{3cm}|p{2cm}|p{2cm}|}
\hline
Approach  & Coverage & Precision \\ \hline
TDL   & 0.0072 & 0.7143 \\ \hline
TDL+ER  & 0.0044 & 0.8235 \\ \hline
TDL+ER+DP  & 0.0036	& 0.8571 \\ \hline
DNN ($\theta$=0.4) &  0.0116 &	0.8444 \\ \hline
DNN ($\theta$=0.7) &  0.0075 &	0.9310  \\ \hline
\end{tabular}
\end{center}
\caption{Precision and coverage of baseline approaches}
\label{tab:baseline}
\vspace{-0.2in}
\end{table}

\begin{figure*}[t]
\vspace{-0.06in}
\begin{minipage}{0.47\linewidth}
\centering
 \includegraphics[width = 3.5in,clip]{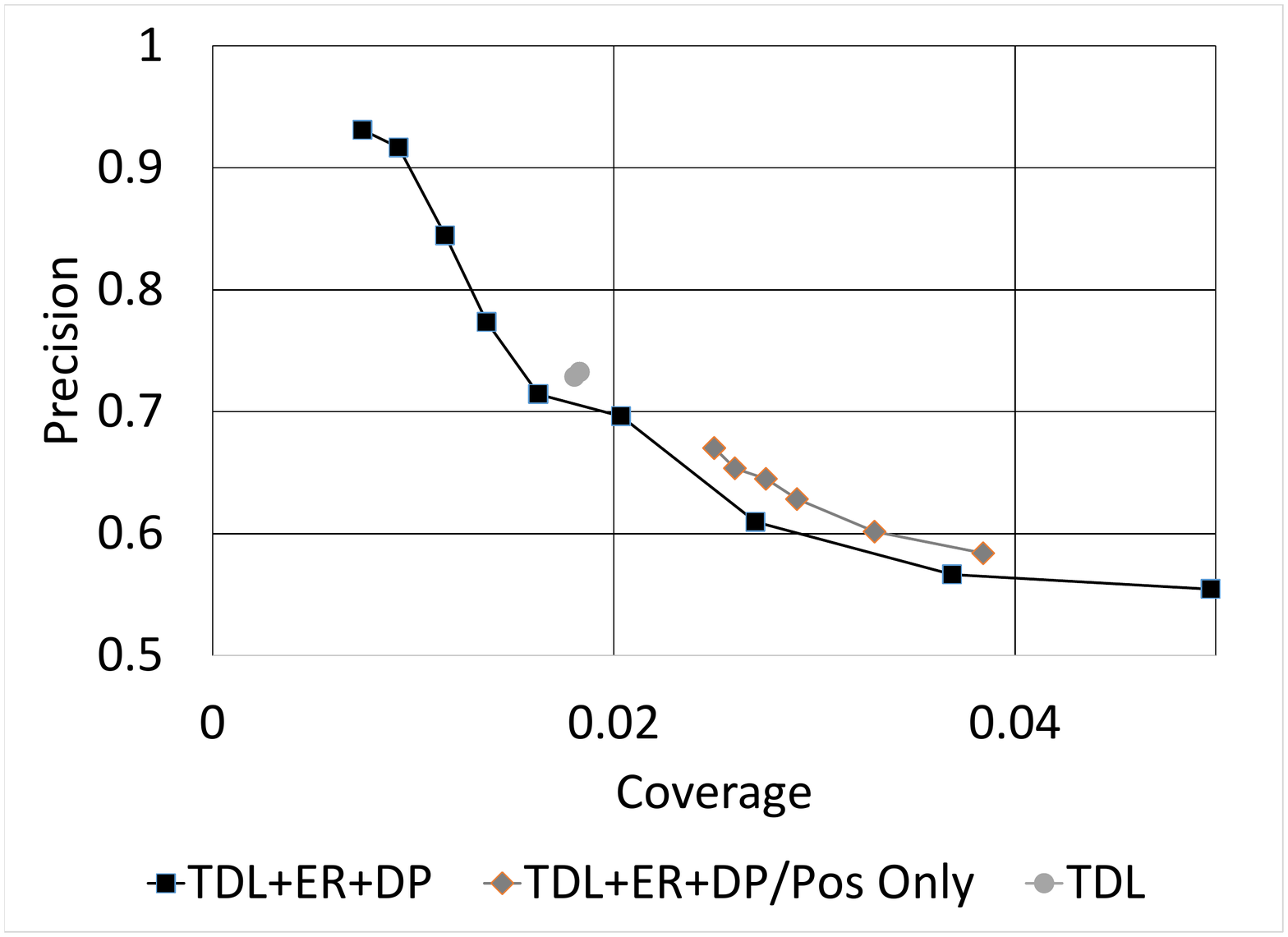}
  \vspace{-0.4in}
 \caption{\small Impact of training data} \label{fig:trainingdata}
\vspace{-0.1in}
\end{minipage}
\hspace{0.05\linewidth}
\begin{minipage}{0.47\linewidth}
\centering
 \includegraphics[width = 3.5in,clip]{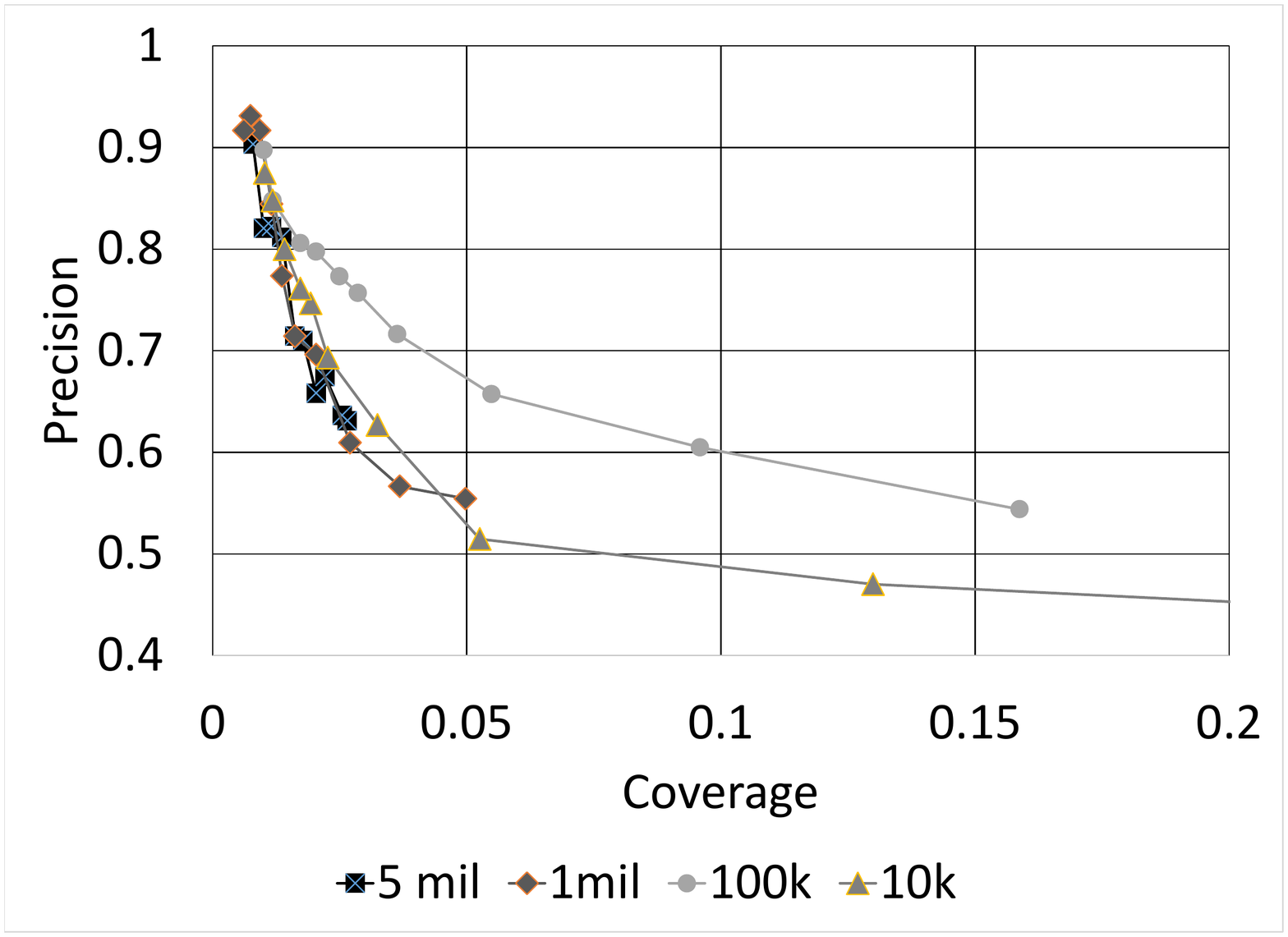}
 \vspace{-0.4in}
 \caption{\small Impact of training data size} \label{fig:trainingdatasize}
\vspace{-0.1in}
\end{minipage}
\end{figure*}

\subsection{Experiments on Assessing Importance of List-intent Queries} \label{sec:expt1}
In this subsection, we conduct experiments to assess the importance of list and superlative intent queries
in the context of web search.

\subsubsection{Dataset}
While there are several labeled datasets for QA of factoid queries, there exists no such dataset
for QA of list-intent and superlative queries \cite{qabenchmark,pasupat:acl2015,berant:emnlp2013}.
We develop such a dataset using real queries from Bing query logs.
We use this dataset for 
assessing the important of list-intent queries
as well as for the experiments on entity type tagging in Section \ref{sec:expt2}.

We first aggregated most recent 2 year query log of Bing and retained
the queries with at least 100 impressions.
We then took a random sample of about 4000 queries from the above query set.
We sent each query to at least 3 experienced human judges using Microsoft's internal crowdsourcing system.
If the query has list or superlative intent, we asked the judge to
label it `Yes' and mark the PST tokens.
Otherwise, she should label it `No'.
We specifically asked the judged to look for queries seeking list of \emph{entities}, not
any list.
For example, a query seeking for a list of instructions/steps (e.g., ‘how to delete files on iphone’)
should be labeled 'No'. This is because relational tables are typically not appropriate for answering
such queries.
Queries of adult or offensive nature should also be labeled `No'.
Once the judged reach an agreement, we save the query, its label (referred to as the true label) and marked the PST tokens.
This resulted in a labeled set of 3879 queries.

\subsubsection{Experimental Results}
Among the 3879 labeled queries, 683 were labeled `Yes'. This indicates $17.6\%$
of all web search queries are of list or superlative intent.
Since the fraction seemed high, we manually examined the positively labeled  queries.
We found that, in spite of our guidelines, judges have occasionally labeled list-but-not-list-of-entities queries
as `Yes'. But the majority of positively labeled queries are true positives, hence we
believe the fraction is still well above $10\%$.
This validates our claim in Section \ref{sec:intro} that this is an important class of queries.

\subsection{Experiments on Entity Type Tagging} \label{sec:expt2}
We conduct experiments to evaluate the proposed DNN-based approach
for entity type tagging
and compare it with baseline approaches (TDL, TDL+ER and TDL+ER+DP)
We evaluate them using two metrics: precision and coverage.
A query is a true positive if (i) the approach returns non-null output
and the true label is positive \emph{and} (ii) they agree on the PST tokens.
It is a false positive if either (i) the approach returns non-null output
but the true label is negative or (ii) the approach returns non-null output and the true label is positive but they do not agree on the PST tokens.
Precision is $\frac{tp}{tp + fp}$
and coverage is $\frac{tp+fp}{3879}$
where $tp$ and $fp$ denote the number of true positives
and false positives respectively.
We measure coverage instead of recall due to the issue with the labels, i.e., some positively labeled
queries are not truly list-intent. This would have lead to inaccurate recall computation
but does not affect coverage computation.
Recall that the DNN approach has the score threshold knob $\rho$;
each choice of $\rho$ yields different precision and coverage numbers.
We use the dataset described in Section \ref{sec:expt1} to conduct these experiments.
\\
\noindent \textbf{Comparison of DNN-based approach with baseline approaches}:
Table \ref{tab:baseline} compares the precision and coverage
of the proposed DNN-based approach and 3 baseline approaches (TDL, TDL+ER and TDL+ER+DP) presented in Section \ref{sec:baseline}.
TDL has a precision of 0.71; entity name removal improves it to 0.82
while the DP-based root word check further improves it to 0.86.
However, the coverage is still low; the coverage of TDL+ER+DP is only 0.0036.
DNN approach has significantly higher coverage for the same precision;
with $\theta=0.4$, the precision is roughly similar to that of TDL+ER+DP (both around 0.85)
but the coverage is \emph{3X higher} (0.0116 vs 0.0036).
At higher values of $\theta$, the DNN approach can achieve
significantly higher precision compared with TDL+ER+DP.
For example, the precision of the DNN approach at $\theta=0.7$ is
0.93 compared with 0.86 of TDL+ER+DP approach.
This shows the superiority of our DNN-based approach
to the baseline approaches.

\begin{figure*}[t]
\vspace{-0.06in}
\begin{minipage}{0.47\linewidth}
\centering
 \includegraphics[width = 3.5in,clip]{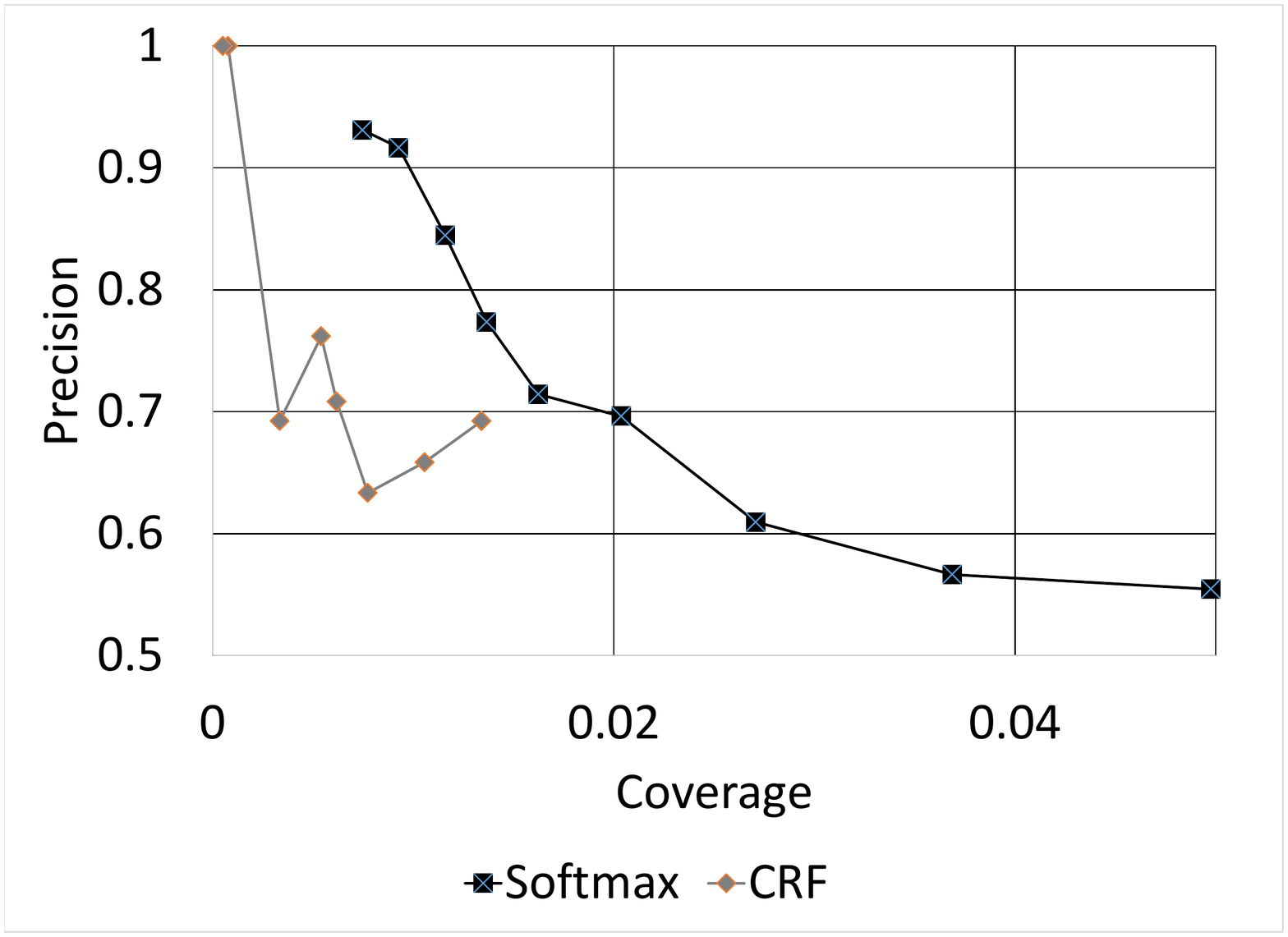}
  \vspace{-0.4in}
 \caption{\small Impact of choice of prediction layer} \label{fig:predictionlayer}
\vspace{-0.1in}
\end{minipage}
\hspace{0.05\linewidth}
\begin{minipage}{0.47\linewidth}
\centering
 \includegraphics[width = 3.5in,clip]{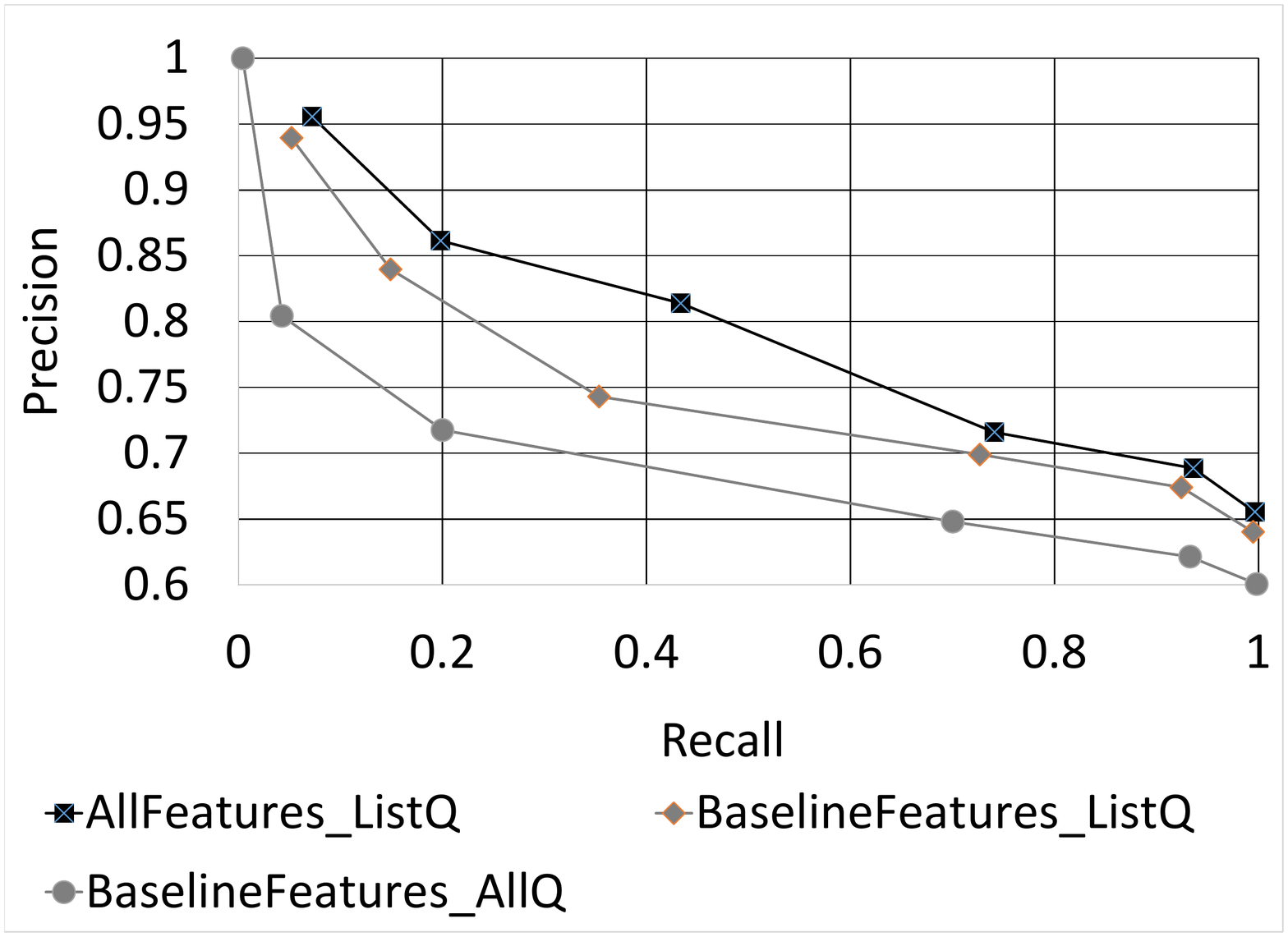}
 \vspace{-0.4in}
 \caption{\small Comparison of table selection approaches for list-intent queries} \label{fig:tableselection}
\vspace{-0.1in}
\end{minipage}
\vspace{-0.1in}
\end{figure*}

\noindent \textbf{Evaluation of training data generation}:
We compare several approaches to automatically generate training data for the DNN model:\\
\noindent $\bullet$ \emph{TDL+ER+DP/BothPosNeg}: This is the proposed approach.
As discussed in Section \ref{sec:tdgen}, we include positive examples, i.e., queries for which TDL+ER+DP produces non null output (50\%)
as well as negative examples, i.e., those discarded by entity name removal (20\%) and DP-based root word check (30\%).
\\
\noindent $\bullet$ \emph{TDL+ER+DP/PosOnly}: We include only positive examples, i.e., queries for which TDL+ER+DP produces non null output.
\\
\noindent $\bullet$ \emph{TDL}: We use the output of TDL approach
as training data.
\\
Figure \ref{fig:trainingdata} compares the precision and coverage of the DNN model
for the above 3 training data generation techniques.
The model trained using TDL+ER+DP/BothPosNeg can reach high precision (in the 80s and 90s)
while the TDL+ER+DP/PosOnly and TDL can reach precision of 73\% and 67\% respectively.
This is because, in the latter two cases, the DNN cannot discriminate between
positive and negative contexts and hence cannot discard the entity name and PST-not-root queries.
Note this was also the problem with TDL (besides having coverage problems) which also has
a precision of 72\% (as shown in Table \ref{tab:baseline}).

\noindent \textbf{Impact of training data size}:
We use \emph{TDL+ER+DP/BothPosNeg} to generate the training data and
vary the total size (taking both positive and negative examples) from 10K to 5 million.
Figure \ref{fig:trainingdatasize} shows the performance of the models
for the various training data sizes.
At the 80-90\% precision levels, all the models perform similarly.
The model trained on 1 million reaches the highest precision (93\% with coverage of 0.0074).
At lower precision levels (below 80\%), the model trainined on 100K outperforms the other models.

\noindent \textbf{Impact of prediction layer}:
We compare two approaches for the final prediction layer in the DNN architecture: CRF and adapted softmax.
Figure \ref{fig:predictionlayer} compares the two approaches.
The softmax approach outperforms the CRF approach although the latter typically
outperforms the former in traditional sequence tagging tasks like named entity recognition and parts-of-speech tagging.
This is because there is at most one non-O tag in a query in our case, so there is no benefit of
holistic prediction (where each prediction depends on neighboring predictions) over local predication.

\subsection{Experiments on Table Answer Selection}

In this subsection, we conduct experiments to evaluate the impact of structure-aware matching features on answer quality.

\subsubsection{Dataset}
We now develop a dataset to evaluate the techniques for table answer selection.
Note that the dataset described in Section \ref{sec:expt1} is for evaluating entity type tagging
(it contained labels for only queries); for evaluating table answer selection, we need
labels for query-table pairs.
As mentioned in Section \ref{sec:expt1}, previously proposed QA datasets
are not suitable for our purpose since they focus
on factoid queries.

We again start with the 2 year aggregated query log described in Section \ref{sec:expt1}.
To create the dataset, we first select a sample of queries.
For each query, we include all the candidate tables for each query.
Recall the candidate tables
is the set of tables occurring in the top $5$ documents returned by the search engine.
Simply taking a random sample of queries with non-empty candidate sets
results in an imbalance in positive and negative examples; there are many
more negative examples than positive ones. This adversely affects the quality of the boosted tree learner we use.
We therefore choose queries where
at least one of the candidate tables dominates the containing document, i.e.,
the table(s)
occupies more than 40\% of the document.
This increases the likelihood of more positive examples
and hence creates a more balanced dataset.
We take a random sample of 15000 queries
from the above set to create our training and evaluation queryset.
This produces 63064 query-table pairs.

We send each pair to at least 3 experienced human judges and ask them to label each query-table pair
as either `Good'(1) or `Bad'(0).
If the table answers the query, is well-formed and comes from a trustworthy source,
the label should be `Good'.
Otherwise, it should be `Bad'.
Once the judges reach an agreement for a pair,
we save the pair and that label (referred to as the true label).
We randomly select 70\% of the queries
for training, 20\% for development and 10\% for test.

\begin{figure*}[t]
\vspace{-0.06in}
\begin{minipage}{0.47\linewidth}
\centering
 \includegraphics[width = 3.5in,clip]{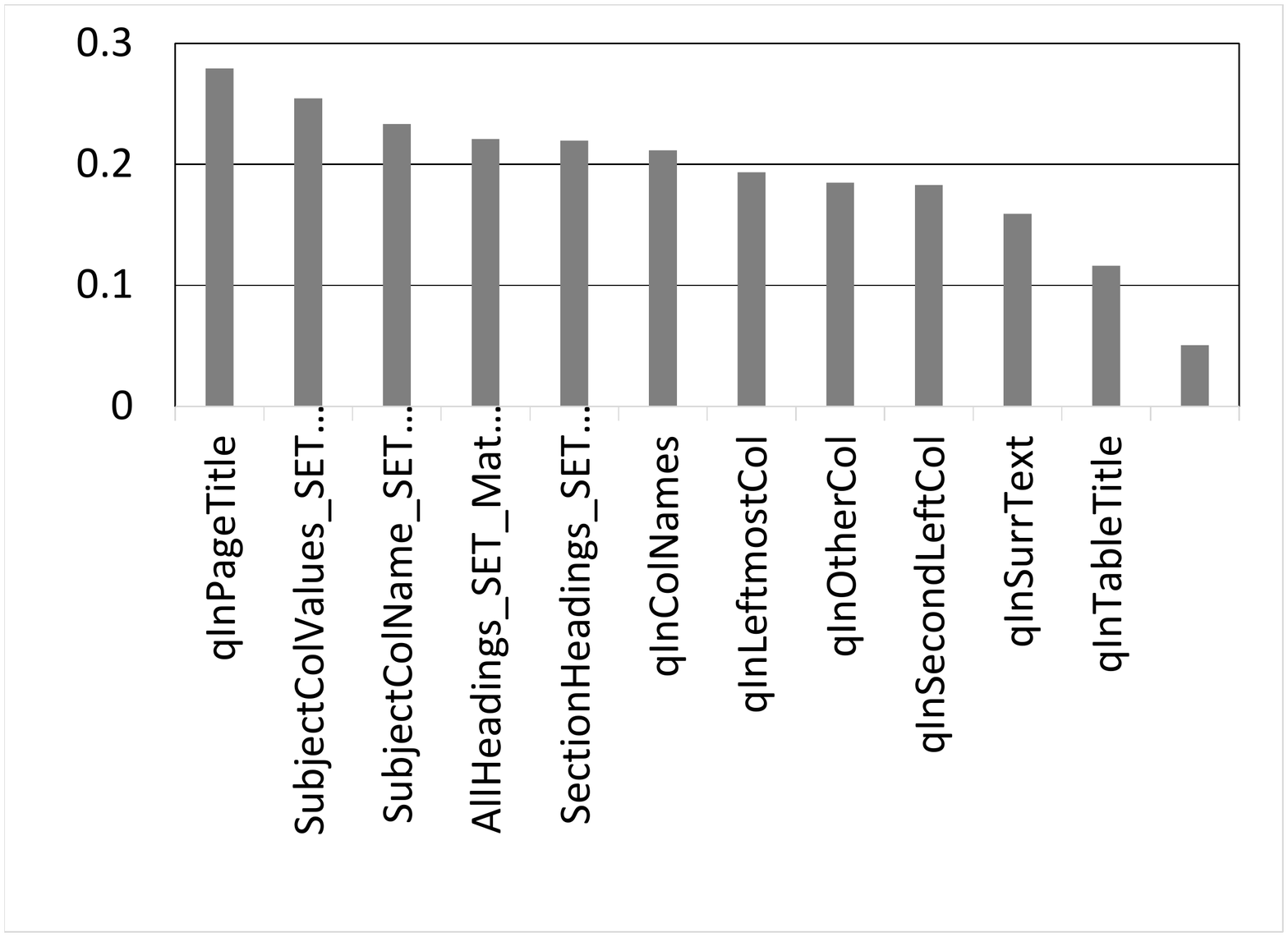}
  \vspace{-0.4in}
 \caption{\small Information gain of table classifier features} \label{fig:featuregain}
\vspace{-0.1in}
\end{minipage}
\hspace{0.05\linewidth}
\begin{minipage}{0.47\linewidth}
\centering
 \includegraphics[width = 3.5in,clip]{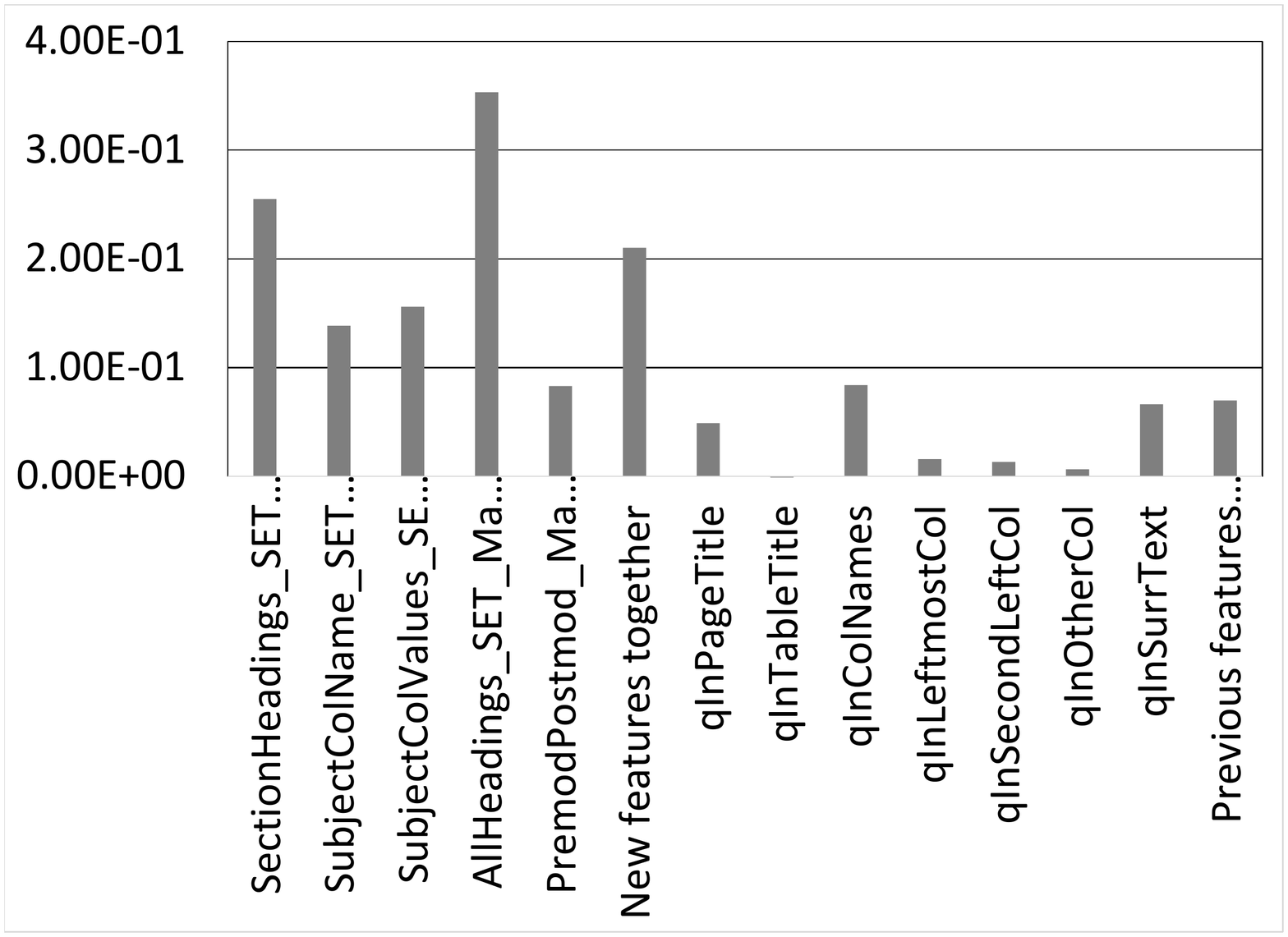}
 \vspace{-0.4in}
 \caption{\small Correlation of features with labels} \label{fig:featurecorr}
\vspace{-0.1in}
\end{minipage}
\end{figure*}

\subsubsection{Experimental results}
\noindent \textbf{Comparison of structure-aware matching with baseline approaches}:
We consider 3 table selection approaches: \\
\noindent $\bullet$ \emph{BaselineFeatures\_AllQ}: This is the baseline approach
that trains the table answer classifier using the features developed for table search (listed in Table \ref{tab:baseline}).
It does not perform any query intent classification, so it performs training and inference for all queries.
\\
\noindent $\bullet$ \emph{BaselineFeatures\_ListQ}: It performs training and inference exactly like the above approach
(i.e., using the features listed in Table \ref{tab:baseline})
but only on the queries for which the entity tagging output is non null.
\\
\noindent $\bullet$ \emph{AllFeatures\_ListQ}: This is the proposed approach that (i) performs
query intent classification/extraction and (ii) uses the structure-aware matching features
along with the features listed in Table \ref{tab:baseline} for table answer selection.
\\

We compare the above approaches in terms of
precision and recall.
For each query, we invoke the table answer selection
algorithm and mark it as one of the following: \\
\noindent $\bullet$ true positive if it returns a result and returned Q-T pair has a true label of 1 \\
\noindent $\bullet$ false positive if it returns a result but returned Q-T pair has a true label of 0 \\
\noindent $\bullet$ false negative if it did not return a result but the query has at least one Q-T pair with true label 1 \\
Precision is $\frac{tp}{tp + fp}$
and recall is $\frac{tp}{tp + fn}$
where $tp$, $fp$ and $fn$ denote the number of true positives,
false positives and false negatives respectively.
Note that we get a precision and recall value for each choice of threshold $\theta$;
we get a precision-recall curve by varying $\theta$.

Figure \ref{fig:tableselection} shows the P-R curve for the 3 table selection approaches.
AllFeatures\_ListQ significantly outperforms the other two approaches.
Since table answers need to have high precision, we operate in the part of the curve
where the precision is between 0.8 and 1.0. For example, \emph{at precision 0.8,
the proposed approach has recall of 0.47 while
BaselineFeatures\_ListQ and
BaselineFeatures\_AllQ have a recall of 0.23 and 0.04 respectively.
At precision 0.9,
the proposed approach has recall of 0.16 while
BaselineFeatures\_ListQ and
BaselineFeatures\_AllQ have a recall of 0.09 and 0.02 respectively}.
This validates that positive impact of entity type tagging
and structure-aware matching on table answer quality.

\eat{
\begin{table}
\begin{center}
\vspace{-0.1in}
\begin{tabular}{|p{5.2cm}|p{1.2cm}|p{1cm}|}
\hline
Feature  & Positive Correlation & Coverage \\ \hline
QT\_NearFieldsMatchTypePlural	&	0.7094	& 0.6625 \\ \hline
QT\_EntityColumnNameHasEntityType	&	0.7169	& 0.3251 \\ \hline
QT\_EntityColumnCellHasEntityType	&	0.7260	& 0.3337 \\ \hline
QT\_MatchTypePlural	&	0.6937	& 0.8557 \\ \hline
Type Match Features Together	&	0.6858	& 0.8887 \\ \hline
NumTokensInPageTitle	&	0.6393	& 0.9621 \\ \hline
NumTokensInTableTitle	&	0.6338	& 0.0550 \\ \hline
NumTokensInAttrNames	&	0.6656	& 0.5452 \\ \hline
NumTokensInLeftmostCol		& 0.6446	& 0.3105 \\ \hline
NumTokensSecondToLeftmostCol	&	0.6430	& 0.3215 \\ \hline
NumTokensInOtherCells		& 0.6388	&	0.3740 \\ \hline
NumTokensInSurroundingText		& 0.6688		&0.3826 \\ \hline
\end{tabular}
\end{center}
\caption{Precision and coverage of baseline approaches}
\label{tab:baseline}
\vspace{-0.2in}
\end{table}
}

\eat{
\begin{figure*}
\vspace{-0.06in}
\centering
 \includegraphics[width = 5.5in,clip]{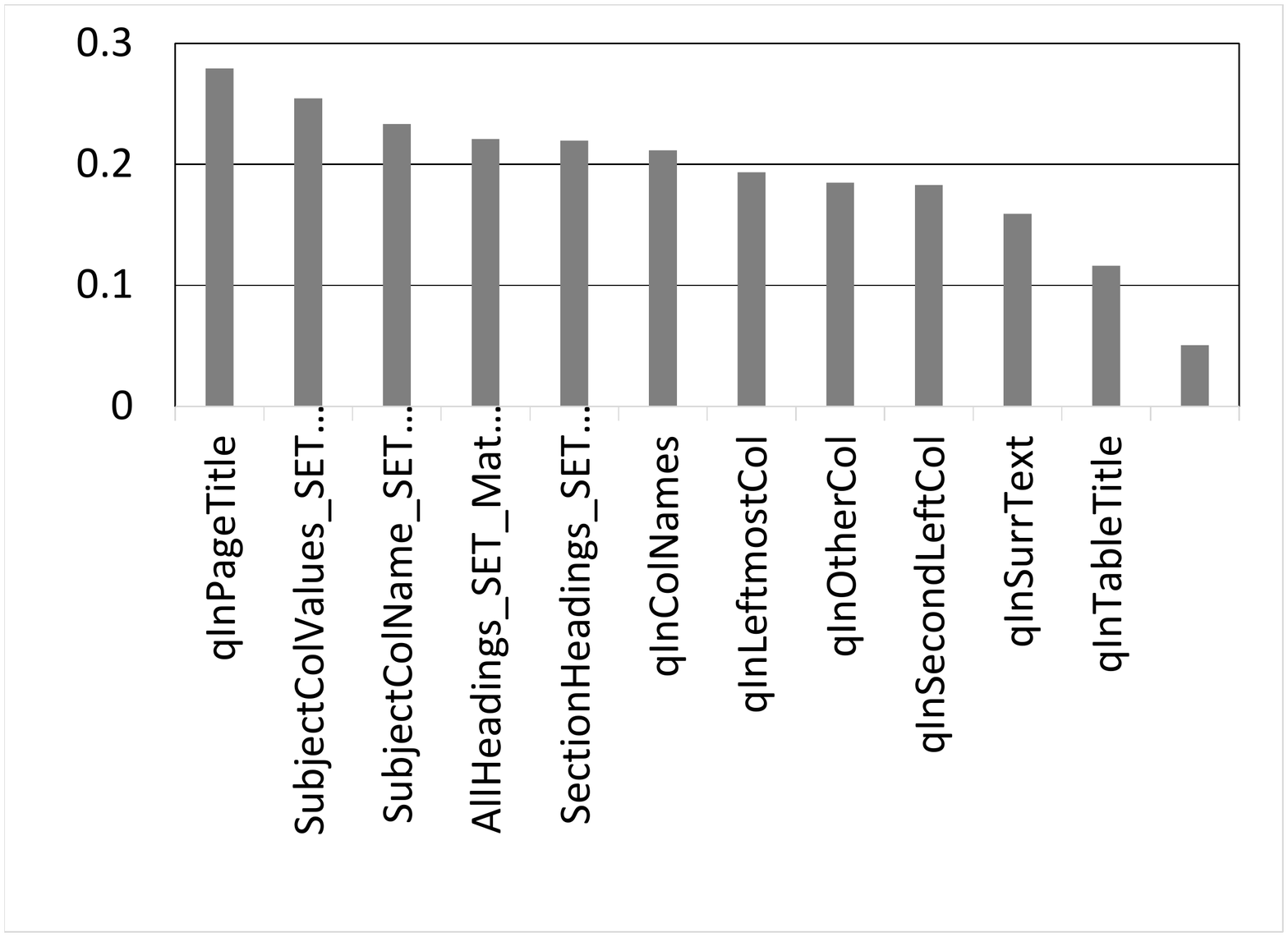}
 \caption{\small Importance of features for table selection} \label{fig:SemanticOverBM25}
  \vspace{-0.15in}
 \end{figure*}
}

\noindent \textbf{Importance of new features}:
We evaluate the importance of the new features introduced for structure-aware match.
Figure \ref{fig:featuregain} shows the information gain for the various Q-T matching features:
the previously proposed text match features as well as the new structure-aware match features.
To focus on comparing the structure-aware match features with textual match features, 
we have removed the table quality and importance features from this chart.
The 4 features that impose match between the SET and type of entities in the table
rank second, third, fourth and fifth in the ranked list of 17 features.
These show the importance of the new features.

While information grain indicates of the importance of the features 
in presence of other features, we plot correlation between the features and the labels
to show the importance of the individual features (or set of features) irrespective of other features.
We compute the correlation using the $\phi$-coefficient which is computed as follows: 
\begin{align*}
\vspace{-0.15in}
\frac{n_{11} \times n_{00} - n_{10} \times n_{01}}{\sqrt{n_{1*} \times n_{0*} \times n_{*0} \times n_{*0}}}
\vspace{-0.15in}
\end{align*}
where $n_{xy}$ denotes the number of test examples with label value $x$ and feature value $y$.
We convert non-boolean features to boolean by setting it to 1 if the feature value is greater than 0
and to 0 otherwise.  
Figure \ref{fig:featurecorr} shows the result.
To compute the correlation for a set of features, 
we compute the feature value by logical OR-ing of the individual feature values.
The leftmost 5 features are the structure-aware match features, they all have much higher correlation
to the labels compared with the textual match features (the ones to the right).
The correlation for the new features together is 0.21 while that of all the previous features together
is 0.0698, i.e., \emph{3X times higher}. This shows the new structure-aware features can predict the labels more accurately
than text match features.

In summary, our experiments show that our proposed framework of intent extraction and structure-aware matching
can produce table answers with higher precision and recall compared with baseline approaches.

\section{Related Work}\label{sec:relwork}

Our work is most related to extensive body of work
on web table search \cite{cafarella:vldb08,venetis:vldb11,bhagavatula:idea13}.
Given a keyword query, the goal is to return a ranked
list of web tables
relevant to the query.
These approaches (specifically features) are not adequate for question answering with tables
as demonstrated by our experiments.
Researchers have studied several search paradigms beyond the above keyword search paradigm:
find related tables \cite{relatedtables_sigmod12}, find tables based on column names \cite{pimplikar:vldb12}
and append new columns to existing entity lists \cite{infogather_sigmod12,zhang:sigmod13}.
However, these paradigms are different from question answering
and hence those techniques cannot be directly applied.

There is also work on question answering using web tables \cite{sun:www2016,facto:www11}.
Both works focus on answering fact lookup queries;
they return the precise cell of a table that answers the question.
Those approaches cannot be easily applied to answer list and superlative intent queries.

Researchers have been studying question answering using text passages for several decades
\cite{pascabook:2003,jurafskybook:qachapter,brill:emnlp2002,lin:tois2007}.
They also focus on fact lookup queries.
They look for textual match between the question with the candidate
passages.
However, as discussed in Section \ref{sec:intro}, textual match alone is not adequate for returning table answers with high precision.
This is validated by our experimental results.

There is a rich body of work on question answering using knowledge bases \cite{berant:emnlp2013,yih:acl2015,fader:kdd2014,unger:www2012,yahya:emnlp2012}.
They focus on fact lookup queries as well.
They parse the question into specific forms such as logic forms, graph queries
and SPARQL queries, which can then be executed again the knowledgebase to find the answer.
These parsing techniques cannot be directly applied for list intent queries.

Our work is also related to works on natural language interfaces to databases \cite{nlidbintro:1995,popescu:iui2003,popescu:coling2004,li:vldb2014,pasupat:acl2015}.
They translate natural language questions to SQL queries which
can then be executed against the database. Like works on question answering using knowledgebases,
they focus on semantic parsing and hence cannot be applied to solve the table answer selection problem.

There is rich body of work on query intent classification into target categories (e.g., Wikipedia categories or concepts)
\cite{queryclass:www09,queryclass:tois06}. These techniques cannot be easily used to detect list and superlative intent.

Our approach to entity type tagging is related to sequence tagging \cite{crf:icml01, huang:2015}. 
Sequence tagging models range from linear statistical models like Hidden Markov models, Maximum entropy
Markov models and CRF \cite{crf:icml01} to neural network based models \cite{huang:2015}.
The latter approaches are the best performing solutions for this task \cite{huang:2015}.
We map our entity tagging task to sequence tagging task
and leverage the best performing solutions developed for it.
\section{Conclusion}
In this paper, we presented \textsc{TableQnA}, a QA system that provides direct answers
to list and superlative intent queries using web tables. Our main insight
is to first extract intent from such queries
and perform structure-aware matching between the extracted intent and the candidates to select the answer.
Our experiments show that our query intent extractor has significantly higher precision
and coverage compared with baseline approaches. Furthermore, by performing structure-aware
matching, our table answer selector outperforms the state-of-the-art baseline.

In future work, we plan to incorporate semantic matching features into the structure-aware match.
For example, we can compute the match between the entity type sought by the query
and the cell values in the subject column of the table in semantic space.
Intent understanding for other classes of queries, e.g., queries seeking for the value of entity on an attribute
and answering them with HTML tables is also an open challenge.

\bibliographystyle{abbrv}
\bibliography{../BibDatabases/references,../BibDatabases/webtablebib,../BibDatabases/nlidb} 

\end{document}